\documentclass[11pt]{amsart}
\usepackage{amsaddr}
\usepackage{graphicx} 
\usepackage{amsmath}
\usepackage{amssymb}
\usepackage{mathpazo}
\usepackage{hyperref}
\usepackage{etoolbox}
\usepackage{setspace}
\usepackage{tikz}
\usepackage{tikz-cd}

\numberwithin{equation}{section}

\newcommand{\bea}{\begin{eqnarray}\displaystyle}
\newcommand{\eea}{\end{eqnarray}}

\newcommand{\del}{\partial}
\newcommand{\ov}{\overline}

\newcommand{\be}{\begin{equation}}
\newcommand{\ee}{\end{equation}}

\newcommand{\IC}{\mathbb{C}}

\usepackage{amsthm}

\newtheorem{theorem}{Theorem}[section] 

\theoremstyle{definition} 

\newtheorem{remark}[theorem]{Remark}

\title{Poisson Vertex Algebras and Three Dimensional Gauge Theory}

\author{Ahsan Z. Khan and Keyou Zeng}

\address{Harvard University,\\
         Center for Mathematical Sciences and Applications,\\
         Cambridge, MA 02138}

\email{\href{mailto:ahsan@cmsa.fas.harvard.edu}{ahsan@cmsa.fas.harvard.edu}, \href{mailto:keyou@cmsa.fas.harvard.edu}{kzeng@cmsa.fas.harvard.edu}}

\date{\today}
\begin{document}

\maketitle

\begin{abstract}
We introduce a mixed holomorphic-topological gauge theory in three dimensions associated to a (freely generated) Poisson vertex algebra. The $\lambda$-bracket of the PVA plays the role of the structure constants of the gauge algebra and the gauge invariance of the theory holds if and only if the $\lambda$-bracket Jacobi identity is satisfied. We show that the holomorphic-topological symmetry of the theory enhances to full topological symmetry if the Poisson vertex algebra contains a Virasoro element. We outline examples associated to PVAs of $\mathcal{W}$-type and demonstrate their connections to various versions of $3d$ gravity. We expect the three-dimensional Poisson sigma model to play an important role in the deformation quantization of Poisson vertex algebras. 
\end{abstract}

\section{Introduction} Vertex algebras play a central role in many areas of mathematical physics, ranging from conformal field theory and string theory to the geometric Langlands program \cite{Frenkel:2005pa}. While vertex algebras remain challenging to study and classify, a much simpler algebraic gadget can be recovered by taking a ``classical limit". Much like the classical limit in quantum mechanics, where the commutator of operators gives rise to a Poisson bracket, the classical limit of a vertex algebra recovers an analogous structure known as a $\lambda$-bracket \cite{li2004vertex,barakat2009poisson}. Roughly speaking, a Poisson vertex algebra (also called Coisson algebra in the language of \cite{beilinsondrinfeld}) is a differential unital commutative algebra equipped with a compatible $\lambda$-bracket. In this note, we show how to construct a three-dimensional gauge theory of a mixed holomorphic-topological nature from a freely generated Poisson vertex algebra. 

Hybrid holomorphic-topological theories generalize the better-studied topological field theories in the sense that they are only topological along some directions but behave holomorphically with respect to the other directions. Such theories are typically more intricate than topological theories but are still well-behaved, often allowing for a rigorous formulation \cite{Wang:2024tjf}. In three dimensions, holomorphic-topological theories were first studied in \cite{Aganagic:2017tvx,Costello:2020ndc} arising as a certain twist of three-dimensional $\mathcal{N}=2$ supersymmetric gauge theories. They were shown to have a rich algebraic structure and to be closely related to $3d$ geometry. In a similar vein, holomorphic-topological twists of $3d$ $\mathcal{N}=4$ theories were studied in \cite{Garner:2022vds,Garner:2023zko}, and exhibit a rich interplay with the story of Coulomb branches and three-dimensional mirror symmetry.

The field theory we introduce is a natural three-dimensional variant of the well-studied Poisson sigma model in two dimensions \cite{Ikeda:1993fh,Schaller:1994es, Cattaneo:1999fm}. It includes all previously mentioned holomorphic-topological twists of $3d$ supersymmetric gauge theories, but also many examples of a different nature. In this paper, we will study examples constructed from Poisson vertex algebras of $\mathcal{W}$-type. It turns out that these theories are closely related to three-dimensional gravity \cite{Collier:2023fwi,Collier:2024mgv,Witten:1988hc}. Their quantization is also closely related to the geometry of the moduli space of Riemann surfaces \cite{Verlinde1990,Kashaev1998}.

Similar to how the two-dimensional Poisson sigma model leads to deformation quantization of Poisson algebras \cite{Kontsevich:1997vb,Cattaneo:1999fm,Cattaneo:2001bp}, we expect that our three dimensional variant will shed light on the problem of deformation quantization of Poisson vertex algebras. More generally, we believe that our model is the correct field theory that generalizes the famous Chern-Simons/WZW correspondence \cite{Witten:1988hf} to more general vertex operator algebras, giving rise to a ``Poisson sigma model/VOA correspondence". Our field theory will thus hopefully prove to be useful in the study of vertex algebras, their representations and their spaces of conformal blocks.

The organization of this note is as follows. We begin in Section \ref{sec:Model} by reviewing some preliminaries of Poisson vertex algebras. We follow this by formulating our model and discussing its gauge transformations and gauge invariance. In Section \ref{sec:Example}, we give several examples of our construction, connecting to Chern-Simons, (super)-gravity, and higher spin theories. In Section \ref{sec:BV}, we reformulate our model in terms of the BV formulation, which proves to be helpful in studying some of its additional properties. In particular, we show that our mixed holomorphic-topological theory is promoted to a fully topological theory if the Poisson vertex algebra contains a Virasoro element. In Section \ref{sec:quant}, we remark on how the model is connected to the problem of deformation quantization. We conclude in Section \ref{conc} and leave some additional remarks to Appendices \ref{super} and \ref{fullyhol}.

\section{Holomorphic-Topological Poisson Sigma Model}
\label{sec:Model}

\subsection{Poisson Vertex Algebras}
There are several equivalent definitions of a Poisson vertex algebra \cite{li2004vertex,barakat2009poisson,beilinsondrinfeld}. In this paper, we adopt the definition that a Poisson vertex algebra $V$ is a unital, commutative, differential algebra equipped with a compatible $\lambda$-bracket. In more detail $V$ is a complex vector space equipped with a commutative product and a derivation $\del: V \rightarrow V$ of this product. If the product of two elements $a,b \in V $ is denoted as $ab$ the operator $\del$ is required to satisfy \be \del(ab) = (\del a)b + a (\del b) .\ee In addition $V$ must carry a $\lambda$-bracket, a map 
\begin{align}
    \begin{split}
        \{\cdot\,_{\lambda} \,\cdot \}: V \otimes V &\rightarrow V[\lambda]\\
        \{a_\lambda b\} &= \sum_{n\geq 0}\frac{\lambda^n}{n!}a_{(n)}b
    \end{split}
\end{align}
subject to the following constraints. 
\begin{description}
    \item[Conformal sesquilinearity] \begin{equation} 
        \{\del a_{\lambda}b \} = -\lambda \{ a_{\lambda} b \}, \quad \{a_{\lambda} \del b \} = (\lambda + \del) \{a_{\lambda} b \}. 
    \end{equation}
    \item[Skew-symmetry] \begin{equation} 
        \{a_{\lambda} b \}= -\,_{\leftarrow}\{b_{-(\del + \lambda)} a \}
    \end{equation}
    \item[Jacobi identity] \begin{equation} \{a \,_{\lambda} \{b \,_{\mu} c\}  \} = \{\{a_{\lambda} b \}_{\lambda + \mu} c\} +  \{b_{\mu} \{a_{\lambda} c\} \}. \end{equation}
   
\end{description} 

In the above it is worth clarifying the skew-symmetry axiom,  where the subscript with the left arrow $\leftarrow$ just means that we move the operator $-(\del + \lambda)$ to the left
\be
\,_{\leftarrow}\{b_{-(\del + \lambda)} a \} = \sum_{n\geq 0}\frac{1}{n!}(-\del - \lambda)^n(b_{(n)}a)\,.
\ee 

The commutative product and the $\lambda$-bracket are required to be compatible in the sense that the left Leibniz rule,
\be \{a_{\lambda} bc \} = \{a_{\lambda} b \} c + \{a_{\lambda}c \} b,\ee is satisfied. 

Skew-symmetry then implies that we also satisfy the right Leibniz rule:
\be 
\{ab_{\lambda} c \} = \{a_{\lambda+ \del} c \}_{\to} b + \{b_{\lambda+ \del} c \}_{\to} a .\ee

In addition to arising as classical limits of vertex algebras, it is also worth noting that Poisson vertex algebras formalize the notion of ``local Poisson brackets" as discussed in the literature on integrable field theories in 1+1 dimensions. One replaces a local Poisson bracket of the form \be \{\phi^i(x), \phi^i(y) \} = \sum_{n} f_n^{ij}(\phi^a, \del_y \phi^a, \del_y^2 \phi^a, \dots) \delta^{(n)}(x-y) \ee where $\delta^{(n)}(x-y)$ denotes the $n$th derivative of a $\delta$-function with the $\lambda$-bracket \be \{\phi^i \,_{\lambda} \, \phi^{j}\} = \sum_{n} f_n^{ij}(\phi^a, \del \phi^a, \dots) (-1)^n \lambda^n. \ee The relationship between PVAs and integrable hierarchies is discussed in more detail in \cite{barakat2009poisson}. 

Let us unpack the type of PVAs we will consider. We take $V$ to be a freely generated $\mathbb{C}[\del]$-module with $N$ generators so that as a commutative algebra we have \be V= \mathbb{C}[\phi^{i}, \del \phi^i, \del^2 \phi^i, \dots]\,,\ee where the index $i$ runs from $1$ to $N$. In this paper, we primarily consider examples where $N$ is finite. However, our theory, at least classically, is also well-formulated for infinite $N$. As shown in \cite{barakat2009poisson}, it is enough to specify the $\lambda$-bracket on the generators and check the Jacobi identity holds. So a Poisson vertex algebra structure is given by \be \{\phi^i \,_{\lambda} \, \phi^j\} = \Pi^{ji}(\lambda),  \ee where $\Pi^{ji}(\lambda)$ is of the form 
\be
\Pi^{ji}(\lambda) = \sum_{n} \Pi^{ji}_n(\phi, \del \phi, \dots) \lambda^n
.\ee The axioms one requires of a $\lambda$-bracket can be expressed in terms of the $\{\Pi^{ji}(\lambda) \}$. Skew commutativity means that that the functions $\{\Pi^{ij} \} $ satisfy 
\be \Pi^{ji}(\lambda) = -\sum_{n} (-\del-\lambda)^n\Pi^{ij}_n(\phi, \del \phi,\dots). 
\ee The $\lambda$-bracket Jacobi identity
\be
\{\phi^i\,_{\lambda} \,\{\phi^j \,_{\mu} \,\phi^k\} \} - \{\phi^j \,_{\mu} \,\{\phi^i \,_{\lambda} \phi^k \} \} = \{\{\phi^i \,_{\lambda} \, \phi^j \} \,_{\lambda+ \mu} \, \phi^k \} 
\ee
becomes  
\begin{align}
\begin{split} \sum_{n \geq 0} \frac{\del \Pi^{kj}(\mu)}{ \del(\del^n \phi^l)} (\del + \lambda)^n \Pi^{li}(\lambda) - \sum_{n \geq 0}\frac{\del \Pi^{ki}(\lambda)}{ \del (\del^n \phi^l)} (\del + \mu)^{n} \Pi^{lj}(\mu)\,\,\,\,\,\,\,\,\,\,\,\,\,\,\,\,\,\,  \\  = \sum_{n \geq 0} \Pi^{kl}(\lambda+ \mu+ \del) (-(\lambda+ \mu + \del))^n \frac{\del \Pi^{ji}(\lambda)}{ \del (\del^n \phi^l)} .\end{split}
\end{align}
Once a collection of $\{\Pi^{ij}(\lambda) \}$ as above are given one can define the Poisson bracket of two arbitrary elements $f, g \in V$ as follows:
\be
\{f \,_{\lambda} \, g\} = \sum_{n,m\geq 0} \frac{\del g}{ \del (\del^n \phi^j)} (\del + \lambda)^n \Pi^{ij} (\del + \lambda) \, (-\del - \lambda)^m \frac{\del f}{ \del (\del^m \phi^i)}.
\ee
An equivalent formulation can be given in terms of differential operators. From the collection of functions $\{\Pi^{ij}(\lambda)\}$ of $\lambda$, we can obtain differential operators simply by replacing a $\lambda^n$ by a $\del^n$ that is meant to act from the left, so that the operator the corresponding differential operator to $\Pi^{ij}(\lambda)$, which we denote as $\Pi^{ij}(\del)$ will be written as 
\be \Pi^{ij}(\del) = \sum_{n \geq 0} \Pi^{ij}_n (\phi, \del \phi, \dots) \del^n.\ee In terms of differential operators the skew-symmetry of the $\lambda$-bracket translates into the condition that $\Pi^{ij}$ is skew-adjoint: so that it satisfies \be \Pi^{ij}(\del) = - \sum_{n \geq 0} (-\del)^n \circ \Pi^{ji}_n(\phi, \del \phi, \dots) = -(\Pi^{ij}(\del))^{\dagger}. \ee The $\lambda$-bracket Jacobi identity can be translated as saying that the differential operators satisfy 
\begin{align} \label{jacobidifferential}
\begin{split}
\sum_{n \geq 0} \Big(\frac{\del \Pi^{kj}(\del)}{ \del (\del^n \phi^l)} G_j\Big)  \big( \del^n \big (\Pi^{li}(\del) F_i \big) \big) - \Big( \frac{\del \Pi^{ki}(\del)}{\del(\del^n \phi^l)} F_i \Big) (\del^n(\Pi^{lj}(\del) G_j) \\ = \sum_{n \geq 0 } \Pi^{kl}(\del) (-\del)^n \Big(G_j \frac{\del \Pi^{ji}(\del)}{ \del (\del^n \phi^l)} F_i \Big). 
\end{split} 
\end{align} Thus the data of a (freely generated) PVA can be equivalently formulated as a collection of matrix valued differential operators satisfying the identities above. For more details on this see Proposition 1.16 (along with equation 1.55) in \cite{barakat2009poisson}. 

We will discuss some standard examples of freely generated Poisson vertex algebras in the following subsection. 

\subsection{The Model}

With these preliminaries out of the way we can begin formulating our model. Let $V$ and $\Pi^{ij}(\lambda)$ be as above. Our field theory is defined on the three-dimensional space $$\mathbb{R}^3 = \mathbb{R} \times \mathbb{C}$$ with standard coordinates $(t, z, \bar{z})$. The fundamental fields of our model consist of $N$ (complex valued) scalar fields \be \phi^i(t,z,\bar{z}), \,\,\,\,\, i=1, \dots, N, \ee along with a collection of $N$ partial one-forms \be \eta_{i} = \eta_{i t}(t,z,\bar{z}) \text{d}t + \eta_{i \bar{z}}(t,z,\bar{z})\text{d} \bar{z}, \,\,\,\, i=1, \dots, N. \ee Let $\text{d}_t + \ov{\del}$ denote the mixed deRham-Dolbeault operator on $\mathbb{R} \times \mathbb{C}$ which acts on the field $\phi$ as \be\label{eqn:diff_HT} (\text{d}_t + \ov{\del}) \phi^i = \frac{\del \phi^i}{\del t} \text{d}t + \frac{\del \phi^i}{ \del\bar{z}} \text{d} \bar{z}. \ee
To define the theory, we choose a holomorphic one-form $\text{d}z$. The action functional of the model is then given by \be \label{eq:non_BV_action} S[\phi, \eta] = \int_{\mathbb{R} \times \mathbb{C}} \text{d}z \Big( \eta_i (\text{d}_t + \ov{\del}) \phi^i + \frac{1}{2} \eta_i \, \Pi^{ij}(\del) \eta_j \Big).   \ee More explicitly the kinetic term is \be\label{eq:act_kin} S_{\text{kin}} = \int \text{d}t \text{d}^2 z \Big(\eta_{i t} \ov{\del}_{\bar{z}} \phi^i - \eta_{i \bar{z}} \del_t \phi^i \Big),\ee whereas the interaction term depending on the PVA structure is given by \be S_{\text{int}} = \frac{1}{2}\int \text{d}z \Big( \sum_{n \geq 0 }\Pi^{ij}_n(\phi, \del \phi, \dots) \eta_i  \del_{z}^n \eta_j \Big). \ee The interaction term can involve holomorphic derivatives of arbitrary order depending on the order of the differential operator $\Pi.$ Our theory is subject to the gauge transformation rules with gauge parameters \be \varepsilon_i(t,z,\bar{z}), \,\,\,\, i=1, \dots, N\ee and gauge transformations 
\begin{align}\label{eqn:general_gauge}\begin{split}
    \delta \phi^i &= \Pi^{ij}(\del) \varepsilon_j , \\ \delta \eta_i &= -(\text{d}_t + \ov{\del}) \varepsilon_i -  \frac{1}{2}\sum_{n \geq 0} (-\del)^n \Big( \eta_j \,\frac{\del \Pi^{jk}(\del)}{\del(\del^n \phi^i)} \varepsilon_k - \varepsilon_j \frac{\del \Pi^{jk}(\del)}{\del (\del^n \phi^i)} \eta_k \Big). 
\end{split}
\end{align}
The first main result of this note is the statement that the action $S$ is invariant under the gauge transformations above if and only if $\Pi^{ij}(\lambda)$ define a Poisson vertex algebra structure on $\mathbb{C}[\phi, \del \phi, \del^2 \phi, \dots]. $

The gauge invariance of $S$ follows from a straightforward calculation. Assuming that $\Pi^{ij}(\del)$ is skew-adjoint, the variation of the action under the gauge transformations up to total derivative terms is 
\begin{align}\label{eq:nBV_g_inv} \begin{split}
\delta S = \int \text{d}z \,\eta_k \Bigg( -\sum_{n\geq 0} \frac{\del \Pi^{kl}(\del)(\varepsilon_l)}{\del (\del^n \phi^i)} \del^n (\Pi^{ij}(\del) \eta_j) + \sum_{n \geq 0} \frac{\del \Pi^{kj}(\del) \eta_j}{ \del (\del^n \phi^i)} \del^n (\Pi^{il} \varepsilon_l) \\ - \sum_{n\geq 0} \Pi^{ki}(\del) (-\del)^n \Big(c_k \frac{\del \Pi^{kl}(\del)}{ \del (\del^n \phi^i)} \eta_l \Big) \Bigg)  .\end{split}
\end{align} We thus see that the variation of the action vanishes if and only if the equation \eqref{jacobidifferential} holds, which is equivalent to the $\lambda$-bracket Jacobi identity. We therefore have a consistent gauge theory which is gauge invariant if and only if the axioms of a Poisson vertex algebra are satisfied. We call it the holomorphic-topological Poisson sigma model. 

One can recast the gauge transformations in BRST form by introducing the ghost fields $c_i$ that correspond to the gauge transformation parameters $\varepsilon_i$. The BRST operator then acts as 
\begin{align}
Q \phi^i &= \Pi^{ij}(\del) c_j , \\ Q \eta_i &= -(\text{d}_t + \ov{\del}) c_i - \frac{1}{2}\sum_{n \geq 0} (-\del)^n \Big(\eta_j \frac{\del \Pi^{jk}(\del)}{ \del(\del^n \phi^i) } c_k - c_j \frac{\del \Pi^{jk}(\del)}{\del (\del^n \phi^i)} \eta_k \Big), \\ Q c^i &= -\frac{1}{2} \sum_{n \geq 0} (-\del)^n \Big( c_j \frac{\del \Pi^{jk}(\del)}{ \del (\del^n \phi^i)} c_k \Big). 
\end{align} The nilpotence of the BRST operator holds if and only if the $\lambda$-bracket Jacobi identity equation is satisfied and the equations of motion 
\begin{align} (\text{d}_t + \ov{\del}) \phi^i + \Pi^{ij}(\del) \eta_j &= 0, \\ -(\text{d}_t + \ov{\del}) \eta_i + \frac{1}{2} \sum_{n \geq 0} (-\del)^n \Big(\eta_i \frac{\del \Pi^{ij}(\del)}{ \del (\del^n \phi^i) } \eta_j \Big) &= 0
\end{align} are obeyed. Later in Section \ref{sec:BV1}, we will reformulate our theory using BV formalism, by further adding anti-fields and anti-ghosts. In the BV formalism, the BRST operator is completed to a BV differential, so that the equation of motion is also forced cohomologically.

\begin{remark}
Since we began with the definition of a Poisson vertex algebra to be a vector space over $\mathbb{C}$ our model is most naturally viewed as an analytically continued theory, namely the action functional is a holomorphic functional on the complex manifold of fields. The phase space similarly will most naturally be viewed as a holomorphic symplectic manifold.
\end{remark}

\subsection{Global definition}
\label{sec:Global_defn}
In this subsection, we study the global definition of our $3d$ theories on more general $3$-manifolds. As discussed in \cite{Aganagic:2017tvx}, a holomorphic-topological theory is not defined on an arbitrary $3$-manifold. Instead, we should consider $3$-manifold equipped with a transverse holomorphic foliation (THF) structure. A THF structure on a three manifold $M$ is given by a covering of coordinate patches with coordinates $(t,z,\bar{z}) \in \mathbb{R}\times \mathbb{C}$, so that the transition
functions between patches take the form
\be
    (t',z',\bar{z}') = (f(t,z,\bar{z}),h(z),\bar{h}(\bar{z})).
\ee
The above form of the transition functions indicates that the $dt$ direction alone is not canonically defined. On the other hand, the space of holomorphic one-forms $\Omega^{1,0}$ is globally well-defined. Using this, we can further define:
\begin{equation}
    \Omega_{\mathrm{HT}}^{\bullet} = \Omega^{\bullet}_{\mathbb{C}}/\Omega^{1,0},
\end{equation}
where $\Omega^{\bullet}_{\mathbb{C}}$ is the complexified de-Rham complex of $M$. The de-Rham differential induces a differential $\mathrm{d}_{\mathrm{HT}}$ on $\Omega_{\mathrm{HT}}^{\bullet}$, which takes the form \eqref{eqn:diff_HT} in a local coordinate patch. 

However, as one can observe from \eqref{eq:non_BV_action}, the definition of our theory further depends on a non-canonical choice of the holomorphic one-form $\text{d}z \in \Omega^{1,0}$. This indicates that not every Poisson sigma model we wrote can be defined on an arbitrary $3$-manifold with a transverse holomorphic foliation structure.\footnote{Strictly speaking, this does not exclude the possibility of defining the theory globally, as it is still possible to find a global section of $\Omega^{(1,0)}$. However, we will not consider this construction, as it would lead to a different theory on a local patch and, consequently, a different boundary vertex algebra, which we will discuss later.} This is to be expected, as (Poisson) vertex algebra is inherently a local notion, and not every Poisson vertex algebra can be defined globally.

As discussed in \cite{frenkel2004vertex}, a (Poisson) vertex algebra can be defined on an arbitrary surface if it admits an action of $\mathrm{Der}(\mathbb{C}[z])$, which is generated by the Virasoro charges $L_{-1},L_0,L_1,\dots$. The action of $L_{-1}$ should be identified with the action of derivative $\del$ of our Poisson vertex algebra. We also require that the action of $L_{0}$ can be diagonalized with half-integer eigenvalues. \footnote{Note that being a Virasoro module allows for rational or even complex weights. The difference here is that a module is only supported at a point on the surface. However, to globalize a vertex algebra, the weights must be restricted to half-integers, as other fractional powers of the canonical bundle do not always exist.} \footnote{There is also a technical assumption that the action of $\mathrm{Der}_+(\mathbb{C}[z]) = \{L_1,L_2,\dots\}$ is locally nilpotent.} 

The action of $\mathrm{Der}(\mathbb{C}[z])$ on a (Poisson) vertex algebra $V$ could be complicated in general. However, we have assumed that $V$ is freely generated. If we further assume that the generators $\phi^i$ are primary fields, the action of $\mathrm{Der}(\mathbb{C}[z])$ can be reduced to the data of an additional ``spin" grading $s$ (usually integral or half-integral) so that $V = \oplus_{s} V^{s}$. For the generators $\phi^i$, the action of $\mathrm{Der}(\mathbb{C}[z])$ is given by
\begin{equation}
\begin{split}
        &L_{-1}\phi^i = \del \phi^i,\quad L_0\phi^i = s_i\phi^i,\\
    &L_n\phi^i = 0,\;\text{ for } n \geq 1.
\end{split}
\end{equation}

The $\lambda$-bracket need to be compatible with the $\mathrm{Der}(\mathbb{C}[z])$ action. This further requires that the $\lambda$-bracket carries spin $-1$, \be \{ \cdot \,_{\lambda} \, \cdot\} : V^{s_1} \otimes V^{s_2} \to (V[\lambda])^{s_1+ s_2 - 1},\ee where we assume the variable $\lambda$ to have spin $1$.

In the case discussed above, one can also formulate the three-dimensional Poisson sigma model on any three-manifold $M$ equipped with a transverse holomorphic foliation. This is done by taking $\phi^i$ to be a section of $(\Omega^{1,0})^{\otimes s_i}$ and $\eta_i$ to be a section of $\Omega_{\mathrm{HT}}^{1}\otimes (\Omega^{1,0})^{\otimes (1- s_i)}$. Locally, they can be written as \be \phi^i = \varphi^i (\text{d}z)^{\otimes s}, \,\,\,\,\, \eta_i = (\eta_{it} \text{d}t + \eta_{i \bar{z}} \text{d} \bar{z} ) \otimes (\text{d}z)^{\otimes (1-s_i)} ,\ee 
When writing the action then \ref{eq:non_BV_action} we no longer require the choice of one-form $\text{d}z.$

However, we will also consider cases where not all the generators are primary fields. In such cases, the $L_0$ action still provides a spin grading $s$, but the action of $\mathrm{Der}(\mathbb{C}[z])$ becomes more complicated. Specifically, a non-primary generator field $\phi^i$ no longer corresponds to a section of $(\Omega^{1,0})^{\otimes s}$ but typically involves a more exotic geometric interpretation. For example, the transformation rule for the Virasoro generator $T$ involves the well-known Schwarzian derivative and is no longer a section of quadratic differentials. We will address these issues on a case-by-case basis later.

\section{Examples}
\label{sec:Example}
In this section, we consider some standard examples of Poisson vertex algebras and their associated holomorphic-topological Poisson sigma models. Some of the examples we discuss also arise from the holomorphic-topological twist of $3d$ $\mathcal{N} = 2$ or $\mathcal{N} = 4$ supersymmetric gauge theories. However there are also theories associated with the Virasoro algebra or $\mathcal{W}$-algebra, which do not have a (known) supersymmetric gauge theory origin. In these cases we find that our examples are closely related to various versions of $3d$ gravity, which is of independent interest.

\subsection{Affine PVA and Chern-Simons Theory} \label{bftheory} The first example we consider is the standard affine Kac-Moody Poisson vertex algebra. Here one has \be \mathcal{V}(\mathfrak{g}) = \mathrm{Sym}(\mathbb{C}[\del] \otimes \mathfrak{g}) \ee for a Lie algebra $\mathfrak{g}$ equipped with a symmetric bilinear form. For a simple Lie algebra, we can let the symmetric bilinear form to be $k\kappa$, where $k\in \mathbb{C}$ and $\kappa$ is the Killing form. By choosing a basis $\{B_a \}$ of $\mathfrak{g}$, we can write the commutative differential algebra as \be \mathcal{V}(\mathfrak{g}) = \mathbb{C}[B_a,\del B_a,\dots] .\ee Then the $\lambda$-bracket is defined by
\be
\{B_a \,_{\lambda} \, B_b \} = f_{ab}^c B_c + k \,\kappa_{ab} \lambda
\ee so that one has 
\be \Pi_{ba}(\del) = f_{ab}^c B_c + k \kappa_{ab} \del .\ee
Thus the holomorphic-topological sigma model has fields given by $B_a$ (renamed from $\phi$) and $A^a_t \text{d}t + A^a_{\bar{z}} \text{d} \bar{z}$ (renamed from $\eta$) with action functional \be S= \int_{\mathbb{R} \times \mathbb{C}}\text{d}z \big(B_a(\text{d}_t + \ov{\del}) A^a + \frac{1}{2} f_{ab}^c B_c A^b A^c + \frac{1}{2} k \,\kappa^{ab} A^a \del A^b \big)   ,\ee
subject to the gauge transformations
\begin{align}
\delta B_a &= -f_{ab}^c B_c \varepsilon^b + k \kappa_{ab} \del \varepsilon^b, \\ \delta A^a  &= -(\text{d}_t + \ov{\del}) \varepsilon^a - f_{bc}^a  \varepsilon^b A^c.
\end{align} When $k =0$ this is the usual holomorphic-topological $BF$ theory as studied in \cite{Gwilliam:2019cbp}. When $k \neq 0$, one can redefine \be \widetilde{A}^a = A^a + \frac{1}{k} \kappa^{ab } B_b \text{d}z ,\ee and under this change of variables $S$ becomes precisely the action of (analytically continued) Chern-Simons theory at level $k$ \cite{Aganagic:2017tvx} \be S = k\int_{\mathbb{R} \times \mathbb{C}} CS(\widetilde{A}).\ee 

We mention that the holomorphic BF/Chern-Simons theory is a special instance of a broader class of theories arising from the holomorphic-topological twist of $3d$ $\mathcal{N} = 2$ supersymmetric theories \cite{Aganagic:2017tvx,Costello:2020ndc}. A $3d$ $\mathcal{N} = 2$ theory can be specified by a Lie algebra $\mathfrak{g}$, a representation $R$ of $\mathfrak{g}$, a Chern-Simons level $k$ and a superpotential $W:R \to \mathbb{C}$. For a vanishing superpotential $W=0$, the theory can be described equivalently by a super Lie algebra
\begin{equation}
    \mathfrak{g}\oplus R[-1]\,.
\end{equation}
The Lie algebra is induced by the Lie bracket on $\mathfrak{g}$ and its action on $R[-1]$\footnote{For a non-vanishing superpotential, this generalizes to an $L_\infty$ algebra structure of $\mathfrak{g} \oplus R[-1]$, but we will not discuss the details here.}. The symmetric bilinear form is still given by the Killing form. The theory is equivalently described as the three-dimensional Poisson sigma model associated to the super Poisson vertex algebra \be V= \text{Sym} \big( \mathbb{C}[\del] \otimes(\mathfrak{g}\oplus R[-1]) \big).\ee Choosing a basis $\{B_a \}$ for $\mathfrak{g}$ and $\{\alpha_i \}$ for $R[-1]$, the $\lambda$-brackets are \begin{align} \{B_a \,_{\lambda} \, B_b \} &= f_{ab}^c B_c + \lambda k \kappa^{ab}, \\ \{B_a \,_{\lambda} \, \alpha_i\} &= \rho(t_a)_{i}^j \alpha_j, \\ \{\alpha_i \,_{\lambda} \,\alpha_j \} &= 0,\end{align} and the corresponding theory is readily seen to be the twisted $\mathcal{N}=2$ theory with gauge algebra $\mathfrak{g}$ and matter representation $R$ with Chern-Simons level $k$.

The holomorphic-topological twist of $3d$ $\mathcal{N} = 4$ theories can be described in the same way \cite{Garner:2022vds}. We can describe it as the holomorphic-topological BF theory associated to the following super Lie algebra:
\begin{equation}
    \mathfrak{g}\oplus \mathfrak{g}^*[-2] \oplus (R^{*}\oplus R)[-1]\,.
\end{equation}
In addition to the Lie bracket induced by the Lie bracket on $\mathfrak{g}$ and the action on $\mathfrak{g}^*[-2] \oplus (R^{*}\oplus R)[-1]$, we further have a Lie bracket induced by the moment map $\mu:R\otimes R^{*} \to \mathfrak{g}$. This Lie algebra is also studied in \cite{Costello:2018swh} and have many applications to the study of the corresponding $3d$ $\mathcal{N} = 4$ theory.

\subsection{Virasoro PVA and Gravity} Our next example concerns the Virasoro Poisson vertex algebra. Let $c\in \mathbb{C}$. The Virasoro PVA $\mathcal{V}ir$ is 
\be 
\mathcal{V}ir = \mathrm{Sym}(\mathbb{C}[\del] T) 
\ee with $\lambda$-bracket given by \be \{T \,_{\lambda} \, T\} = \del T + 2 T \lambda + \frac{c}{12} \lambda^3,\ee so that one has \be \Pi^{TT}(\del) = \del T + 2T \del + \frac{c}{12} \del^3. \ee Before formulating the 3d model, we note that $T$ is a spin-two field. By abuse of notation, we use the same $T$ to denote the 3d field as well. We can write the fundamental fields as:
\begin{align}
T &= T_{zz}(t,z,\bar{z}) \text{d}z^{\otimes 2}, \\ \mu &= \Big(\mu^z_{t}(t,z,\bar{z}) \text{d} t + \mu^z_{\bar{z}}(t,z,\bar{z}) \text{d}\bar{z} \Big) \otimes \frac{\del}{\del z} \,.
\end{align} The action functional is given by 
\be S = \int_{\mathbb{R} \times \mathbb{C}} \mu (\text{d}_t + \ov{\del}) T + T \mu \del \mu + \frac{c}{24} \mu \del^3 \mu 
\ee subject to the gauge transformations 
\begin{align}
\label{eqn:gauge_T}\delta T &= \frac{c}{12} \del^3 \varepsilon + 2 T \del \varepsilon + \del T \varepsilon, \\ \delta \mu &= -(\text{d}_t + \ov{\del}) \varepsilon - 
\mu \del \varepsilon + \varepsilon \del \mu.
\end{align} 
In the above formulation, we treat $T$ as a section of $(\Omega^{1,0})^{\otimes 2}$. This identification is valid for $c = 0$ but becomes non-canonical when $c \neq 0$. In fact, the gauge transformation rule \eqref{eqn:gauge_T}, when restricted to holomorphic functions, corresponds to the infinitesimal form of the transformation of the stress tensor under a coordinate change $w = \varphi(z)$:
\begin{equation}
    \widetilde{T}(\varphi(z)) = \left(\frac{\partial \varphi}{\partial z}\right)^2 T(\varphi(z)) - \frac{c}{12}\{\varphi;z\},
\end{equation}
where $\{\varphi;z\}$ is the Schwarzian derivative
\begin{equation}
    \{\varphi;z\} = \frac{\partial^3_z\varphi}{\partial_z\varphi} - \frac{3}{2}\left(\frac{\partial^2_z\varphi}{\partial_z\varphi}\right)^2\,.
\end{equation}
We briefly review several equivalent interpretations of the above formula \cite{frenkel2004vertex,Frenkel:2005pa}. The first interpretation is that $T(z)$ is an element of the space of projective connection, which is an affine space modeled on quadratic differential. By definition, a projective connection is a second order differential operator $D:K^{-\frac{1}{2}} \to K^{\frac{3}{2}}$ \footnote{Here we only describe the algebraic version of the construction. The $3d$ fields are smooth, and their description is more subtle, but it reduces to the algebraic description after we impose the equations of motion.} such that the principal symbol is $1$ and the subprincipal symbol is $0$. Explicitly, a projective connection can be written as
\begin{equation}
    \partial_z^2 + u(z)
\end{equation}
in a local coordinate. For $c \neq 0$, one can check that $\partial_z^2 + \frac{6}{c}T(z)$ transform exactly as a projective connection.

The second interpretation is more geometric in nature. We can identify a projective connection with a projective structure. A projective structure on a curve can be understood as an equivalence class of local coordinate charts, where the transition functions between overlapping charts are given by Möbius transformations.
\begin{equation}
    z \to \frac{az+b}{cz+d}\,.
\end{equation}
We refer to \cite{frenkel2004vertex} for the proof of the bijection between the sets of projective connections and projective structures. This interpretation is also useful for understanding the solutions to the equations of motion of the classical theory. As we mentioned in Section \ref{sec:Global_defn}, our theory can be defined globally on a three-manifold with a THF structure. Then the field $\mu$ is a deformation of this THF structure, and the field $T$ further promotes the THF structure so that it is transversely projective.

The third interpretation identifies $T(z)$ as an $\mathfrak{sl}_2$-oper. Roughly speaking, an oper is a (equivalent class of) distinguished connection on a principal bundle. In the $\mathfrak{sl}_2$-case, a oper can be represented by the following connection
\begin{equation}
    \partial_z + \begin{pmatrix}
        0&u(z)\\1&0
    \end{pmatrix}\,.
\end{equation}
After a coordinate transformation, the connection will acquire non-zero diagonal entries. But one can verify that it can be brought back to the above form via a gauge transformation valued in the upper triangular matrices.  The transformation rule for $u$ then takes the same form as that for the projective connection. While we will not discuss this in detail here, the oper interpretation provides the most natural framework for generalization to other $\mathcal{W}$-algebras.

\subsubsection{Phase space and Quantization}
As a $3d$ theory, our model associates a Hilbert space with a Riemann surface $\Sigma$. For holomorphic topological theories with a $3d$ $\mathcal{N} = 2$ origin, the corresponding Hilbert space is carefully analyzed in \cite{Bullimore:2018yyb}. Notably, this theory that correspond to Virasoro algebra does not arise from any $3d$ $\mathcal{N} = 2$ gauge theory. Instead, we take the approach outlined in \cite{Costello:2020ndc} by analyzing the phase space of this model. We begin by writing down the equations of motion:
\begin{align} (\text{d}_t + \ov{\del}) T  - \del T \mu  -2T\del\mu+ \frac{c}{12} \del^3 \mu = 0, \\ -(\text{d}_t + \ov{\del})\mu  + \mu \del \mu = 0.  \end{align} In a local patch of the form $\Sigma\times[-\delta,\delta]$, we can choose the gauge with  $\mu_t^z = 0$. We still denote $\mu = \mu_{\bar{z}}^zd\bar{z}\frac{\partial}{\partial z}$. When $c$ is set to zero, the equations of motion simplify to:
\begin{equation}
    \begin{aligned}
        &d_t\mu = d_tT = 0\,,\\
        &(\bar{\partial} + \mathcal{L}_{\mu})T = 0\,,\\
    \end{aligned}
\end{equation}
where $\mathcal{L}_{\mu}$ is the Lie derivative with respect to $\mu$. The gauge transformation for $\mu$ take the following form
\be\label{eqn:bel_gauge}
\delta\mu = -\bar{\del}\varepsilon - \mu\del\varepsilon + \varepsilon\partial\mu \,.
\ee
Recall that a Beltrami differential also transforms as a $T^{1,0}_{\Sigma}$-valued $(0,1)$-form under holomorphic coordinate transformations. Under a diffeomorphism $\varphi$, the Beltrami differential transforms as 
\begin{equation}
\mu \to \frac{\bar{\partial}\varphi + \mu(\varphi) \bar{\partial}\bar{\varphi}}{\partial \varphi + \mu(\varphi) \partial \varphi}\,.
\end{equation}
For an infinitesimal diffeomorphism $\varphi(z, \bar{z}) = z + v(z, \bar{z})$, the above formula reproduces \eqref{eqn:bel_gauge} if we set $\varepsilon = v + \mu \bar{v}$. Hence, the gauge transformation for $\mu$ corresponds to the infinitesimal action of the diffeomorphism group $\mathrm{Diff}_{0}(\Sigma)$\footnote{We denote $\mathrm{Diff}(\Sigma)$ the space of orientation preserving diffeomorphism of $\Sigma$ and $\mathrm{Diff}_0(\Sigma)$ those diffeomorphisms that are homotopic to identity.}, and we can identify $\mu = \mu_{\bar{z}}^z(z, \bar{z})d\bar{z} \frac{\partial}{\partial z}$ as a Beltrami differential. For $\Sigma$ a genus $g$ Riemann surface, it is known that the quotient space
\begin{equation}
    \{\text{Beltrami differential on } \Sigma\}/\mathrm{Diff}_{0}(\Sigma)
\end{equation}
is the Teichm\"{u}ller space $\mathcal{T}_g$. Notice that the Teichmüller space has an action of the mapping class group $ \mathrm{MCG} = \mathrm{Diff}(\Sigma)/\mathrm{Diff}_0(\Sigma)$, which is not generated via infinitesimal gauge transformations. The quotient $\mathcal{T}_g / \mathrm{MCG}$ is known as the moduli space of Riemann surfaces $\mathcal{M}_g $. One can also choose $\mathrm{Diff}(\Sigma)$ as the gauge group instead $\mathrm{Diff}_0(\Sigma)$, which produce $\mathcal{M}_g$ in the phase space. It is merely a subtle aspect of the definition of our theory whether we obtain $\mathcal{T}_g$ or $\mathcal{M}_g$.

The remaining equations for $T$ simply require it to be a holomorphic section of quadratic differential. In fact, the holomorphic quadratic differential parametrize the cotangent bundle of $\mathcal{T}_g$. To see this, note
that the tangent space of $\mathcal{T}_g$ is given by $H^1(\Sigma,T_{\Sigma})$. We then use Serre duality
\begin{equation}
    H^{1}(\Sigma,T_{\Sigma})^{*} = H^{0}(\Sigma,K_{\Sigma}^{\otimes 2})\,.
\end{equation}
To summarize, we find that the space of solutions to the equations of motion on a surface of genus $g$ is the total space of the cotangent bundle
\begin{equation}
    T^*\mathcal{T}_g
\end{equation}
of the Teichm\"{u}ller space $\mathcal{T}_g$. As this phase space should be understood as a holomorphic symplectic manifold, we use a holomorphic version of the geometric quantization. Then the Hilbert space is given by holomorphic sections of certain line bundle on $\mathcal{T}_g$. Though we have used $c = 0$ in our analysis, a non-zero $c$ only twist this line bundle by the determinant line bundle on the Teichm\"{u}ller space. We emphasis that the central charge will be shifted at the quantum level. This can be understood as the metaplectic correction in geometric quantization. We will also provide a derivation of this correction in Section \ref{sec:q_correction}. This Hilbert space is also known to give rise to the Virasoro conformal block. It is also related to a three-dimensional gravity in some recent proposal \cite{Collier:2023fwi,Collier:2024mgv}. A more detailed analysis of the Virasoro Poisson sigma model and its relationship with $3d$ gravity will appear in a separate work.

\subsubsection{Wilson-type Line Defects} We can construct Wilson line operators for this theory. We aim to construct Wilson lines labeled by $\mathfrak{sl}_2$ representations generated by a highest weight vector. It will be useful to regard $\mathfrak{sl}_2$ as the Lie subalgebra of 
\be
\mathfrak{witt}_+ = \{\ell_{n},n\geq -1\} \,,
\ee where the commutation relation is 
\be 
[\ell_n,\ell_m] = (n-m)\ell_{n+m}.
\ee
We identify $\mathfrak{sl}_2$ as the subalgebra $\{\ell_{-1},\ell_0,\ell_1\}$. Suppose we have a irreducible representation $(R,\rho)$ of $\mathfrak{sl}_2$ with highest weight vector $v_h$. It is easy to extend it to a representation of $\mathfrak{witt}_+$ by imposing $\rho(\ell_n)v_h = 0$ for $n \geq 1$. Then, we define the Wilson line operator by 
\begin{equation}
    \mathbb{W}_{R} = P\exp\left(\int_{\mathbb{R}_t}\rho(\tilde{\mu}_t)\right),
\end{equation}
where
\begin{equation}
   \tilde{\mu}_t =  \sum_{n\geq 0}\frac{1}{n!}\partial^n\mu_t \ell_{n-1}\,.
\end{equation}
To see that $\mathbb{W}_{R}$ is indeed a valid Wilson line operator, we notice that the gauge transformation for $\mu_t$ is equivalent to
\begin{equation}
    \delta(\frac{1}{n!}\partial^n\mu_t) =  -\frac{1}{n!} \text{d}_t \partial^n \varepsilon + \sum_{k = 0}^{n+1}\frac{(n - 2k + 1)}{k!(n-k+1)!}\partial^k\varepsilon \partial^{n-k+1}\mu_t,\quad n\geq 0.
\end{equation}
We can further bring it to the following expression
\begin{equation}
    \delta\tilde{\mu}_t = - d_t\tilde{\varepsilon} - [\tilde{\varepsilon},\tilde{\mu}_t]
\end{equation}
where 
\be 
\tilde{\varepsilon} = \sum_{n\geq 0}\frac{1}{n!}\partial^n\varepsilon \,\ell_{n-1}. 
\ee
Then the gauge invariance property of $\mathbb{W}_{R}$ follows the standard argument as in ordinary Chern-Simons theory.

\subsection{SuperVirasoro and Supergravity} \label{sugra} Our construction works equally well for Poisson vertex superalgebras. In this section, we consider the $\mathcal{N}=1$ super Virasoro algebra. There are two super extensions of the Virasoro algebra, known as the Ramond algebra and the Neveu-Schwarz algebra. Here, we focus on the Neveu-Schwarz algebra. It has an even generator $T$, as before, and an odd generator $G$, with the $\lambda$-brackets:
 \begin{align}
     \{T \,_{\lambda} \, T \} &= \frac{c}{12} \lambda^3 + 2T \lambda + \del T, \\ \{T \,_{\lambda} \, G \} &= \frac{3}{2} G \lambda + \del G, \\ \{G \,_{\lambda} \, G\} &= \frac{c}{3} \lambda^2 + 2T. 
 \end{align} 
The corresponding Poisson sigma model has a field content consisting of bosonic fields $T$ and $\mu$, as before, along with their superpartners:
\begin{align}
    G &= G (\text{d}z)^{\otimes \frac{3}{2}}, \\ \psi &= \big(\psi_t \text{d}t + \psi_{\bar{z}} \text{d} \ov{z} \big) \otimes \sqrt{\frac{\del}{\del z} }.
\end{align}
They are fermionic fields of spin $\frac{3}{2}$ and spin $-\frac{1}{2}$ respectively. The action functional is
\begin{align}
     S = \int_{\mathbb{C} \times \mathbb{R}} \Big( & \mu (\text{d}_t + \ov{\del} ) T  + \psi(\text{d}_t + \ov{\del} \big) G +  \frac{c}{24} \mu \del^3 \mu + T  \mu \del \mu \nonumber \\ & \,\,\,\,\,\,\,\,\,\,\,\,\,+ \frac{3}{2} G  \psi \del \mu + \del G   \psi \mu- \frac{c}{6} \psi \del^2 \psi -  T \, \psi \psi \Big).
\end{align} From the general prescription \ref{eqn:general_gauge}, the gauge transformation contains both an even part and an odd part. Let the corresponding even and odd gauge parameters be given by $\varepsilon$ and $\gamma$, respectively. The gauge transformation acts on the fields $T,G$ by
\begin{align}\label{eqn:sVir_gauge1}
\begin{split}
 \delta T &= \frac{\mathrm{c}}{12} \del^3 \varepsilon + 2 T \del \varepsilon + \del T \varepsilon - \frac{1}{2} \del G \gamma - \frac{3}{2} G \del \gamma, \\ 
\delta G &= -\del G \varepsilon - \frac{3}{2} G \del \varepsilon - \frac{\mathrm{c}}{3} \del^2 \gamma - 2T \gamma  \,. 
\end{split}
\end{align} On the one-form fields it acts via 
\begin{align}\label{eqn:sVir_gauge2}\begin{split}
        \delta \mu &= -(\text{d}_t + \ov{\del}) \varepsilon -\mu \del \varepsilon +\varepsilon \del \mu + 2 \gamma \psi , \\ 
        \delta \psi &= -(\text{d}_t+ \ov{\del})\gamma -\del \psi \,\varepsilon + \frac{1}{2} \psi \del \varepsilon - \del \gamma \, \mu  + \frac{1}{2} \gamma \del \mu.  
\end{split}
\end{align} 
The gauge transformation parametrized by $\gamma$ is noteworthy. By setting $\varepsilon = 0 $ and $\gamma$ to be a constant, we find
\begin{align}
        \delta T & = -\frac{1}{2}\gamma \del G,\quad\delta G = -2\gamma T,\\ 
        \delta \mu &= 2\gamma \psi,\quad \delta \psi = \frac{1}{2}\gamma\del \mu.
\end{align}
We observe that this gauge symmetry extends the supersymmetry of the super-Virasoro algebra, given by $\delta T = -\frac{1}{2} \partial G$ and $\delta G = -2T$. This is a distinctive feature of supergravity, where supersymmetry is promoted to a local gauge symmetry. However, it is important to note that the supersymmetry mentioned above is not the usual $3d$ supersymmetry, as the $3d$ Poincaré symmetry is broken in the holomorphic topological theory.

As in the Virasoro case, treating $T$ as a quadratic differential is not natural for $c \neq 0$. In this case, we can combine $T,G$  and define a super field
\begin{equation}
    \mathfrak{T} = G + \theta T\,.
\end{equation}
The gauge transformation \ref{eqn:sVir_gauge1} can be written as
\begin{equation}\label{eqn:gauge_super}
    \delta\mathfrak{T} = (v\partial_z + \frac{1}{2}(D_{\theta}v)D_{\theta} + \frac{3}{2}\partial_zv)\mathfrak{T} + \frac{c}{12}D_{\theta}\partial_z^2v
\end{equation}
where $$ D_\theta := \partial_{\theta} + \theta\partial_z$$ and $v(z,\theta) = \varepsilon + 4\theta\gamma$. We show in  Appendix \ref{app:Super_Proj} that $\mathfrak{T}$ form a super projective connection. As studied in \cite{Zeitlin:2013iya}, super projective connections also have equivalent interpretations as super projective structures, or as $\mathfrak{osp}(1|2)$-oper. We discuss this in more detail in Appendix \ref{app:Super_Proj}.

\subsubsection{Classical phase space}
Again, we analyze the classical phase space of this model and its quantization. In fact, the geometry of the phase space is closely related to the moduli space of super Riemann surfaces. We provide a brief review in Appendix \ref{app:Super_RS} and refer to \cite{Witten:2012ga,Donagi:2013dua,Friedan:1986rx} for more details. We begin by formulating the equations of motion:
\begin{align}
    &(\text{d}_t +\bar{\del})\mu - \mu\del\mu + \psi\psi = 0\,,\\
    &(\text{d}_t +\bar{\del})\psi + \frac{1}{2}\psi\del\mu - \del\psi \mu = 0\,,\\
    &(\text{d}_t + \ov{\del}) T  + \del T \mu  + 2T\del\mu+ \frac{c}{12} \del^3 \mu + \frac{1}{2}\del G\psi + \frac{3}{2}G\del\psi= 0\,,\\
    &(\text{d}_t + \ov{\del}) G + \frac{3}{2}G\del\mu + \del G \mu + \frac{c}{3}\del^2\psi + 2T\psi = 0\,.
\end{align}
Note that we have two gauge parameters $\varepsilon,\gamma$, so that we can choose the gauge such that both $\mu_t$ and $\psi_t$ vanish. To simplify the discussion, we focus on the case when $c = 0$. The equations of motion simplify if we use the super field $\mathfrak{T} = G + \theta T$. We also review in Appendix \ref{app:Super_RS} that the pair
\be (\mu , \psi)  \in \Omega^{0,1}(\Sigma,T_\Sigma\oplus T^{\frac{1}{2}}_{\Sigma}) \ee 
can be understood as a $(0,1)$ form valued in the sheaf $\mathcal{S}$ of super-conformal vector fields of the super Riemann surface $\mathfrak{X}$:
\begin{equation}
     \mathfrak{S}\frac{\partial}{\partial z} + \frac{D_{\theta}\mathfrak{S}}{2}D_{\theta}\,,
\end{equation}
where $\mathfrak{S} = (\mu_{\bar{z}} + 4\theta\psi_{\bar{z}})d\bar{z}$. We can write the equations of motion as
\begin{align}
    &\text{d}_t\mathfrak{S} = 0,\quad \text{d}_t \mathfrak{T} = 0\,,\\
    &\label{eqn:EOM_sT}(\bar{\del} + \mathcal{L}_{\mathfrak{S}})\mathfrak{T} = 0\,.
\end{align}
Here, the Lie derivative $\mathcal{L}_{\mathfrak{S}}$ of $\mathfrak{T}$ take the standard form  \cite{Friedan:1986rx}
\begin{equation}
        \mathcal{L}_{\mathfrak{S}}\mathfrak{T} = (\mathfrak{S}\partial_z + \frac{1}{2}(D_{\theta}\mathfrak{S})D_{\theta} + \frac{3}{2}\partial_z\mathfrak{S})\mathfrak{T}\,.
\end{equation}
According to \cite{Witten:2012ga}, $H^1(\mathfrak{X},\mathcal{S})$ parametrizes the deformations of super Riemann structures. There it is found that one can interpret the deformation as a deformation of the antiholomorphic derivative $\bar{\del}$ by
\begin{align}
    \begin{split}
            \bar{\partial}' &= \bar{\partial} + \mathfrak{S}\frac{\partial}{\partial z} + \frac{D_{\theta}\mathfrak{S}}{2}D_{\theta}\\
            &= \bar{\partial} + (\mu_{\bar{z}} + \frac{1}{2}\mu_{\bar{z}}\theta\partial_{\theta}) + \psi_{\bar{z}} (\partial_{\theta} - \theta\partial)\,.
    \end{split}
\end{align}

The pair of fields $(G, T)$ can be understood as sections of $K^{\otimes 2} \oplus K^{\otimes \frac{3}{2}}[1]$. In Appendix \ref{app:Super_RS}, we show that the superfield $\mathfrak{T} = G + \theta T$ can be interpreted as a smooth section of a bundle $\mathcal{D}^{-3}$ of the super Riemann surface. The equation of motion \eqref{eqn:EOM_sT} simply means that $\mathfrak{T}$ is a holomorphic section under the deformed complex structure $\bar{\partial}'$. Furthermore, as analyzed in Appendix \ref{app:Super_RS}, $H^0(\mathfrak{X},\mathcal{D}^{-3})$ parametrizes the cotangent bundle of the supermoduli space via a super version of the Serre duality.

To summarize, we find that the phase space that our theory assigns to a surface is the total space of the cotangent bundle to the super Teichm\"{u}ller space or super moduli space. Then the Hilbert space is given by the space of holomorphic sections of the super Teichm\"{u}ller or moduli space.

\subsubsection{Wilson-type Line Defects}
We can also construct the Wilson line operators analogous to the Virasoro model. First, we introduce the super Witt algebra $\mathfrak{switt}$. It is defined as the Lie superalgebra of superconformal vector fields on the punctured super-disk $\mathring{D}^{1|1}$. In local coordinates $(z,\theta)$, superconformal vector fields are vector fields that preserve $D_{\theta} = \frac{\partial}{\partial \theta} + \theta \frac{\partial}{\partial z}.$ We review in Appendix \ref{app:Super_RS} that they can be written as
\begin{equation}
    f(z,\theta)\frac{\partial}{\partial z} + \frac{D_{\theta}f(z,\theta)}{2}D_{\theta}\,.
\end{equation}
We define the following basis
\begin{align}
        \ell_n &= z^{n+1}\frac{\partial}{\partial z} + \frac{n+1}{2}z^{n}\theta\frac{\partial}{\partial\theta},\quad n \in \mathbb{Z},\\
        \mathrm{g}_r &= z^{r+\frac{1}{2}}(\frac{\partial}{\partial \theta} - \theta \frac{\partial }{\partial z}),\quad r \in \mathbb{Z}+ \frac{1}{2}\,.
\end{align}
They have the commutation relation
\begin{equation}
    [\ell_n,\ell_m] = (n-m)\ell_{n+m},\quad [\ell_n,\mathrm{g}_r] = (\frac{n}{2} -r)g_{n+r},\quad [\mathrm{g}_r,\mathrm{g}_s] = 2\ell_{r+s}\,.
\end{equation}
The super Witt algebra $\mathfrak{switt}$ contains a subalgebra $\mathfrak{switt}_+$, spanned by the basis elements $\{\ell_n \mid n \geq -1\}$ and $\{\mathrm{g}_r \mid r \geq -\frac{1}{2}\}$. Within $\mathfrak{switt}_+$, there exists a further subalgebra spanned by the basis $\{\ell_{-1}, \ell_0, \ell_1, \mathrm{g}_{-\frac{1}{2}}, \mathrm{g}_{\frac{1}{2}}\}$. In fact, this subalgebra can be identified with $\mathfrak{osp}(1|2)$.

We define 
\begin{equation}
    \tilde{\boldsymbol{\mu}} =\sum_{n\geq 0} \frac{1}{n!}\partial^n\mu_t\ell_{n-1} + \frac{1}{n!}\partial^n\psi_t\mathrm{g}_{n-\frac{1}{2}}\,.
\end{equation}
Then we can check that the gauge transformation \eqref{eqn:sVir_gauge2} for $\mu_t,\psi_t$ is equivalent to the following expression
\begin{equation}\label{eq:par_gauge_sVir}
    \delta\tilde{\boldsymbol{\mu}} = \mathrm{d}_t\tilde{\boldsymbol{\varepsilon}} + [\tilde{\boldsymbol{\varepsilon}},\tilde{\boldsymbol{\mu}}]\,,
\end{equation}
where the $\tilde{\boldsymbol{\varepsilon}}$ is defined as
\begin{equation}
    \tilde{\boldsymbol{\varepsilon}} = \sum_{n\geq 0}\frac{1}{n!}\partial^n\varepsilon \,\ell_{n-1} + \frac{1}{n!}\partial^n\gamma \,\mathrm{g}_{n-\frac{1}{2}}\,.
\end{equation}
Therefore, we can define Wilson line operator associated to representation of $\mathfrak{osp}(1|2)$ as follows. Let $(R,\rho)$ be a irreducible representation of $\mathfrak{osp}(1|2)$ with highest weight vector $v_h$. We can extend it to a representation of $\mathfrak{switt}_+$ by imposing the conditions $\rho(\ell_n)v_h = 0$ for $n \geq 1$ and $\rho(\mathrm{g}_r)v_h = 0$ for $n \geq \frac{1}{2}$. We can define the following Wilson line operator
\begin{equation}
    \mathbb{W}_{R} = P\exp\left(\int_{\mathbb{R}_t}\rho(\sum_{n\geq 0} \frac{1}{n!}\partial^n\mu_t\ell_{n-1} + \frac{1}{n!}\partial^n\psi_t\mathrm{g}_{n-\frac{1}{2}})\right)\,.
\end{equation}
As before, the gauge invariance property of $\mathbb{W}_{R}$ is guaranteed by \eqref{eq:par_gauge_sVir}.

\subsection{$W_3$ PVA and Higher Spin Theory} A more complicated and perhaps less familiar example can be obtained by considering model corresponding to the Poisson vertex algebra arising as the classical limit of the $W_3$-vertex algebra. It can be written as follows. It is generated by two fields $T$ of spin $2$ and $W$ of spin $3$ \be V = \mathrm{Sym}(\mathbb{C}[\del] T \oplus \mathbb{C}[\del]W)\ee with the $\lambda$-brackets 
\begin{align} 
   \{ T \,_{\lambda}\,T \} &= \del T + 2T \lambda + \frac{c}{12} \lambda^3, \\
    \{T \,_{\lambda} \, W \} &= \del W + 3 W \lambda, \\
    \{W \,_{\lambda} \, W \} &= \frac{c}{360} \lambda^5 + \frac{1}{3} T \lambda^3 + \frac{1}{2} \del T \lambda^2 + \Big(\frac{3}{10} \del^2 T + \frac{32}{5c} \Big)\lambda + \frac{1}{15} \del^3 T + \frac{32}{5c} T \del T.
\end{align} The corresponding HT Poisson sigma model has fields $(T,\mu)$ as in the Virasoro theory, and in addition 
\begin{align}
W &= W_{zzz}(\text{d}z)^{\otimes 3}, \\ \chi &= (\chi_t \text{d}t + \chi_{\bar{z}} \text{d}\bar{z}) \otimes (\frac{\del}{\del z})^{\otimes 2}.
\end{align} The corresponding action functional takes the form \be S = S_{\text{kin}} + S_{TT} + S_{TW} + S_{WW}\ee where 
\begin{align}
    S_{\text{kin}} = \int  \mu(\text{d}_t + \ov{\del}) T+  \chi (\text{d}_t + \ov{\del}) W
\end{align} 
is the usual kinetic term and the other three terms are the interaction terms coming from the different non-trivial $\lambda$-brackets as follows \begin{align} S_{TT} &= \int T \mu \del \mu + \frac{c}{24} \mu \del^3 \mu, \\ S_{TW} &= \int 3 W \chi \del \mu + \del W \chi \mu  , \\ 
 S_{WW} &= \int \frac{c}{360} \chi \del^5 \chi + \frac{1}{3} T \chi \del^3 \chi + \frac{1}{2} \del T \chi \del^2 \chi + \frac{3}{10} \del^2 T \chi \del \chi + \frac{32}{5c} T^2 \chi \del \chi.\end{align} 
We have two gauge transformation parameters $\varepsilon$ and $\theta$ which now act on the fields as follows. On the $(T,W)$-fields we have
\begin{align}
\delta T &= \frac{c}{12} \del^3 \varepsilon + \del T \varepsilon + 2T \del \varepsilon + 2 \del W \theta + 2W \del \theta, \\
\delta W &= \del W \varepsilon + 3 W \del \varepsilon + \frac{c}{360} \del^5 \theta + \frac{1}{3} T \del^3 \theta 
+ \frac{1}{2} \del T \del^2 \theta \notag \\
&\quad + \bigg(\frac{3}{10} \del^2 T + \frac{32}{5c} T \bigg) \del \theta 
+ \frac{1}{15} \del^3 T \, \theta + \frac{32}{5c} T \del T \theta\,.
\end{align} 
The action of the gauge transformations on the one-form fields $\mu$ and $\chi$ reads 
\begin{align}
\delta \mu &= -(\text{d}_t + \ov{\del}) \varepsilon - \mu \del \varepsilon + \varepsilon \del \mu 
- \frac{1}{15} \chi \del^3 \theta + \frac{1}{15} \theta \del^3 \chi \notag \\
&\quad + \frac{1}{10} \del \chi \del^2 \theta - \frac{1}{10} \del \theta \del^2 \chi 
- \frac{32}{5c} T (\chi \del \theta - \theta \del \chi), \\
\delta \chi &= -(\text{d}_t + \ov{\del}) \theta - \mu \del \theta + \varepsilon \del \chi 
+ 2 \del \mu \theta - 2 \del \varepsilon \chi.
\end{align}

The natural geometric interpretation of the fields $(T,W)$ is as an $\mathfrak{sl}_3$-oper. To see this, we recall that the space of $\mathfrak{sl}_3$-opers can be identified with the space connections taking the form
\begin{equation}
    \del_z + \begin{pmatrix}
        *&*&*\\+&*&*\\0&+&*
    \end{pmatrix},
\end{equation}
modulo gauge transformation consisting of upper triangular matrices. Here $*$ indicates an arbitrary function and $+$ indicates a nowhere vanishing function. It can be shown that each such oper may be represented by a unique connection of the form
\begin{equation}
    \del_z + p_{-1} + u_1p_1 + u_2p_2
\end{equation}
where \be p_{-1} = \begin{pmatrix}
    0&0&0\\1&0&0\\0&1&0
\end{pmatrix},  \,\,\,\,\,\, p_1 = \begin{pmatrix}
    0&2&0\\0&0&2\\0&0&0
\end{pmatrix}, \ee and \be p_2 = p_1^2 .\ee One can show that under changes of coordinates, $u_1$ transforms as a projective connection (in our notation $T$) and $u_2$ transforms as a holomorphic cubic differential (in our notation $W$) \cite{Frenkel:2002fw}.

We can also consider opers for $\mathfrak{sl}_n$ and other simple Lie algebras \cite{Drinfeld1985}. They correspond to classical $\mathcal{W}_n$ algebras and other $\mathcal{W}$ algebras from classical Drinfeld-Sokolov reduction. They also have natural Poisson vertex algebra structures that allow us to define the $3d$ theories. It would be interesting to give a precise relation between these field theories and higher spin gravities.

\section{Further properties}
\label{sec:BV}
\subsection{BV formulation}\label{sec:BV1} Since our theory involves complicated and nonlinear gauge transformations in general, it will be useful to analyze them using the Batalin–Vilkovisky (BV) formalism. The BV formalism encodes the gauge structure, equations of motion, and symmetry compactly into cohomological data via the BV-BRST differential. In addition to the ghost $ c_i $ that corresponds to the gauge transformations, we introduce anti-fields $ \eta^{i+}, \phi_i^{+} $, and the anti-ghost $ c^{i+} $. As in \cite{Costello:2020ndc}, the entire BV field content can be concisely packaged into superfields. To achieve this, we recall that in Section \ref{sec:Global_defn}, we defined the complex $(\Omega_{\mathrm{HT}}^{\bullet} = \Omega^{\bullet}_{\mathbb{C}} / \Omega^{1,0}, \text{d}_{\mathrm{HT}})$. In a local coordinate patch $ (t, z, \bar{z})$, $\Omega_{\mathrm{HT}}^{\bullet}$ can be identified with $C^{\infty}(\mathbb{R}^3)[\text{d}t, \text{d}\bar{z}]$. More explicitly, we have
\begin{equation}
    \Omega_{\mathrm{HT}}^{0} = C^{\infty}(\mathbb{R}^3),\quad \Omega_{\mathrm{HT}}^{1} = C^{\infty}(\mathbb{R}^3) \text{d}t\oplus C^{\infty}(\mathbb{R}^3) \text{d}\bar{z},\quad \Omega_{\mathrm{HT}}^{2} = C^{\infty}(\mathbb{R}^3) \text{d}t \text{d}\bar{z}\,.
\end{equation}
The differential is given by 
\be
\text{d}_{\mathrm{HT}} = \text{d}_t + \bar{\del} = \text{d}t\frac{\partial}{\partial t} + \text{d}\bar{z}\frac{\partial}{\partial \bar{z}} .
\ee
Let $L = \mathbb{C}^N$. Using $\Omega_{\mathrm{HT}}^{\bullet}$, we can write the superfields as
\begin{align}
    &\mathbf{\Phi}^i = \phi^i+ \eta^{i+} + c^{i+} \in \Omega_{\mathrm{HT}}^{\bullet}\otimes L\,,\\
    &\boldsymbol{\eta}_i = c_i + \eta_i + \phi_i^+ \in \Omega_{\mathrm{HT}}^{\bullet}\otimes L^\vee[1]\,.
\end{align}
We have the canonical BV bracket, which is a shifted Poisson bracket:
\begin{equation}
    \{\mathbf{\Phi}^i(x),\boldsymbol{\eta}_j(y)\}_{\text{BV}} = \delta^i_j\delta(x-y) \text{d} t \text{d}\bar{z}\,.
\end{equation}
The BV action functional is easy to write down. We simply replace the fields by the BV superfields in the action \ref{eq:non_BV_action}. As before, we need to use the holomorphic one form $\text{d}z \in \Omega^{(1,0)}$. We have
\begin{equation}\label{eqn:BV_action}
    S_{\text{BV}} = \int_{\mathbb{R} \times \mathbb{C}} \text{d}z\Big( \boldsymbol{\eta}_i (\text{d}_t + \bar{\del}) \mathbf{\Phi}^i + \frac{1}{2}\sum_{n\geq 0} \boldsymbol{\eta}_i  \Pi_n^{ij}(\mathbf{\Phi},\partial \mathbf{\Phi}^i,\dots) \partial^n \boldsymbol{\eta}_j \Big).
\end{equation}
The BV-BRST differential is easy to compute from the BV bracket, via the formula \be Q_{\text{BV}} = \{ S_{\text{BV}},-\}_{\text{BV}} .\ee More explicitly, we have 
\begin{align}
    &Q_{_{\text{BV}}}\mathbf{\Phi}^i = (\text{d}_t + \bar{\del})\mathbf{\Phi}^i + \Pi^{ij}(\partial)\boldsymbol{\eta}_j\,,\\
    &Q_{_{\text{BV}}}\boldsymbol{\eta}_i = (\text{d}_t + \bar{\del})\boldsymbol{\eta}_i+ \frac{1}{2}\sum_{n\geq 0}(-\partial)^n\boldsymbol{\eta}_j  \frac{\partial \Pi^{jk}(\partial)}{\partial(\partial^n\mathbf{\Phi}^i)} \boldsymbol{\eta}_k\,.
\end{align}
The consistency of this theory is given by the classical master equation \be Q_{_{\text{BV}}}^2 = \{ S_{\text{BV}},S_{\text{BV}}\}_{\text{BV}} = 0.\ee We can check that this is guaranteed by the Jacobi identity of the $\lambda$-bracket, just as we proved the gauge invariance of the action \ref{eq:nBV_g_inv}.

As discussed in Section \ref{sec:Global_defn}, the theory is only defined locally on $\mathbb{C} \times \mathbb{R}$, as the holomorphic one-form $\text{d}z \in \Omega^{(1,0)}$ depend on the choice of coordinates. To define it globally, additional conditions on the Poisson vertex algebra are required, such as the existence of an action of $\mathrm{Der}(\mathbb{C}[z])$. For simplicity, we focus on the case where all the generators $\phi^i \in L$ are primary fields. We decompose $L$ according to their spin $L = \oplus_{s}L^s$. For convenience, we define the auxiliary complex
\begin{equation}
    \Omega_{\mathrm{HT}}^{\bullet,(k)} = \Omega_{\mathrm{HT}}^{\bullet}\otimes(\Omega^{(1,0)})^{\otimes k}
\end{equation}
which is also equipped with the differential $\text{d}_{\mathrm{HT}}$. In a local coordinate patch, $\Omega_{\mathrm{HT}}^{\bullet,(k)}$ takes the form $C^{\infty}(\mathbb{R}^3)[\text{d}t, \text{d}\bar{z}]\text{d}z^{k}$. Using  $\Omega_{\mathrm{HT}}^{\bullet,(k)}$ , we can write the superfields as
\begin{align}
    &\mathbf{\Phi}^i \in \Omega_{\mathrm{HT}}^{\bullet,(s_i)}\otimes L^{s_i}\,,\\
    &\boldsymbol{\eta}_i \in \Omega_{\mathrm{HT}}^{\bullet,(1-s_i)}\otimes (L^{s_i})^\vee[1]\,.
\end{align}
The BV action functional takes the same form as \ref{eqn:BV_action}, except that $\mathrm{d}z$ is now absorbed into the definition of the fields.

\begin{remark}
    Given the success of AKSZ sigma models \cite{Alexandrov:1995kv} as a tool to organize the ordinary two-dimensional Poisson sigma model \cite{Cattaneo:2001ys}, it is natural to wonder if the three-dimensional Poisson sigma model also has such a formulation. This is indeed so, however the AKSZ model has to be formulated with an infinite-dimensional supermanifold as the target space. In more detail, we first define a $QP$-manifold that controls Poisson vertex algebras, analogous to the shifted cotangent bundle of a Poisson manifold $X$ which controls ordinary Poisson structures. The directions of this manifold consist of fields $\boldsymbol{\varphi}^i, \boldsymbol{c}_i$ which are anti-holomorphic forms on $\mathbb{C}$ that are valued in $L$. The symplectic form is given by \be \omega = \int_{\mathbb{C}} \text{d}z \, \delta \boldsymbol{c}_i \wedge \delta \boldsymbol{\varphi}^i.\ee The homological vector field $Q$ acts on these via 
    \begin{align} Q \boldsymbol{\varphi}^i &= \ov{\del} \boldsymbol{\varphi}^i + \Pi^{ij}(\del) \boldsymbol{c}_j, \\ Q \boldsymbol{c}_i &= -\ov{\del} \boldsymbol{c}^i + \sum_{n}(-\del)^n \Big(\boldsymbol{c}_j \frac{ \del \Pi^{jk}(\del) }{\del( \del^n \boldsymbol{\phi}^i)} \boldsymbol{c}_k \Big),
    \end{align} which is square zero if and only if the Jacobi identity holds. The 3d Poisson sigma model is then the AKSZ theory with this target space, with the domain space being the supermanifold underlying the deRham complex on $\mathbb{R}$.
\end{remark}

\subsection{Gauge Algebra from Poisson Vertex Algebra} 
The geometry of the classical BV formalism can be summarized as a differential graded (shifted)-symplectic structure on the space of BV fields. Letting $\mathbf{I}$ be a local functional, it induces a Hamiltonian vector field \be X_{\mathbf{I}} = \{\int \mathbf{I},-\}_{\mathrm{BV}} .\ee Now supposing that $X_{\mathbf{I}}$ preserves the BV action functional $S_{\mathrm{BV}}$, or equivalently \be Q_{\mathrm{BV}} \mathbf{I} = 0, \ee we have that $X_{\mathbf{I}}$ generates a symmetry of the classical system.\footnote{We emphasize that the condition is modified for the quantum theory \cite{Sen:1993ic,Schwarz:1993xp}.} One can also understand this symmetry as a deformation of the BV differential/BV action functional $S \to S'$ where \be S' = S_{\text{BV}}+ \int \mathbf{I} \ee so that $S'$ satisfy the classical master equation \be \{S',S'\}_{\mathrm{BV}} = 0 .\ee

There is an obvious way to construct local functionals that satisfy $Q_{\mathrm{BV}} \mathbf{I} = 0$, by letting $\mathbf{I} = Q_{\mathrm{BV}}\mathbf{\Lambda}$. The corresponding Hamiltonian vector field $\{\int Q_{\mathrm{BV}}\mathbf{\Lambda},-\}_{\mathrm{BV}}$ should be understood as a gauge symmetry. Typically, they are considered trivial and are modded out in the BRST-BV complex. In this section, we will analyze a special class of gauge symmetries in the three-dimensional Poisson sigma model.

Given any element $\Lambda(\phi^i,\del\phi^i,\dots)$ in our Poisson vertex algebra $V$, we can promote it into a superfield as follows
\begin{equation}
    \mathbf{\Lambda} = \Lambda(\mathbf{\Phi}^i,\del\mathbf{\Phi}^i,\dots)\,.
\end{equation}
We find that
\begin{equation}
    Q_{\mathrm{BV}}\mathbf{\Lambda} = (\mathrm{d}_t + \bar{\del})\mathbf{\Lambda} + \sum_{l \geq 0 } \frac{\partial\mathbf{\Lambda}}{\partial( \partial^l\mathbf{\Phi}^k)}\partial^l(\mathbf{\Pi}^{kj}(\partial)\boldsymbol{\eta}_j)\,.
\end{equation}
We can simplify $\int \text{d}z \, Q_{\mathrm{BV}}\mathbf{\Lambda}$ by using the skew-symmetry of $\lambda$ bracket and integration by parts, which gives us
\begin{equation}
    \int \text{d}z \, Q_{\mathrm{BV}}\mathbf{\Lambda} = \int \text{d}z \sum \{\mathbf{\Lambda}_{\lambda}\mathbf{\Phi}^j\}|_{\lambda = 0}\boldsymbol{\eta}_j\,.
\end{equation}
Let us denote \be \delta_{\Lambda} := \{\int \text{d}z\, Q_{\mathrm{BV}}\mathbf{\Lambda},-\}_{\mathrm{BV}}. \ee $\delta_{\Lambda}$ then acts on the fields as 
\begin{align}
    &\delta_{\Lambda}\mathbf{\Phi}^i = -\{\mathbf{\Lambda}_\lambda\mathbf{\Phi}^i\}|_{\lambda = 0}\,,\\
    &\delta_{\Lambda}\boldsymbol{\eta}_i =  \frac{\delta}{\delta \mathbf{\Phi}^i}\left(\int dz \{\mathbf{\Lambda}_{\lambda}\mathbf{\Phi}^j\}|_{\lambda = 0}\boldsymbol{\eta}_j\right)\,.
\end{align}

Let us compute the algebra of gauge transformation generated by $\Lambda \in V$. Commutator of gauge transformation can be computed as \cite{Sen:1993ic} 
\begin{equation}
    \delta_{\Lambda_1}\delta_{\Lambda_2} - \delta_{\Lambda_2}\delta_{\Lambda_1} = \{Q_{\mathrm{BV}}\frac{1}{2}\left(\{\int \mathbf{\Lambda}_1,Q_{\mathrm{BV}}\int\mathbf{\Lambda}_2\}_{\mathrm{BV}} - \{\int \mathbf{\Lambda}_2,Q_{\mathrm{BV}}\int\mathbf{\Lambda}_1\}_{\mathrm{BV}} \right),-\}_{\mathrm{BV}}\,.
\end{equation}
We can check that
\begin{equation}
\frac{1}{2}\left(\{\int \mathbf{\Lambda}_1,Q_{\mathrm{BV}}\int\mathbf{\Lambda}_2\}_{\mathrm{BV}} - \{\int \mathbf{\Lambda}_2,Q_{\mathrm{BV}}\int\mathbf{\Lambda}_1\}_{\mathrm{BV}} \right) = \{\mathbf{\Lambda}_1~_{\lambda}\mathbf{\Lambda}_2\}|_{\lambda = 0}\,.
\end{equation}
Therefore, the Lie bracket $[\Lambda_1, \Lambda_2]$ defiend by $ \delta_{[\Lambda_1,\Lambda_2]} = \delta_{\Lambda_1}\delta_{\Lambda_2} - \delta_{\Lambda_2}\delta_{\Lambda_1} $ coincides with the $\lambda$-bracket of $\{\Lambda_1 \,_{\lambda} \Lambda_2 \}$ evaluated at $\lambda=0,$ namely \be [\Lambda_1, \Lambda_2] = \{\Lambda_1 \,_{\lambda} \Lambda_2 \}|_{\lambda = 0}. \ee  We must also be careful to remember that $\Lambda$ enters the gauge transformation formula only after integration. Therefore if $\Lambda$ is a total $\del$-derivative $\Lambda = \del U$ for some $U \in V$ then one automatically has $\int\del U = 0$. Therefore, we can identify the gauge algebra as the Lie algebra:
\begin{equation}
\left(V/\del V,\{~_{\lambda}\}|_{\lambda = 0}\right)\,.
\end{equation}
This Lie algebra acts on the Poisson vertex algebra $V$, and also plays an important role in the theory of Hamiltonian PDE \cite{barakat2009poisson}. It is an important feature that this symmetry of the vertex Poisson algebra becomes gauge symmetry in the corresponding $3d$ sigma model. We have seen a special case of this phenomenon in Section \ref{sugra}, where the supersymmetry becomes a gauge symmetry. We analyze another important corollary of this fact in the next section.

\subsection{Virasoro Algebra and Topological Theory}In this section, we point out an important property of the three-dimensional Poisson sigma model. As formulated above, the model is holomorphic in the $z$-direction but topological only in the $\mathbb{R}$-direction. However, often the gauge symmetry enhances so that the theory becomes fully topological. The simplest example of this can be seen by looking at the theory corresponding to the affine Poisson vertex algebra with $k\neq 0$. As noted in Section \ref{bftheory} this is equivalent to standard Chern-Simons theory at level $k$, which is indeed fully topological. A wider class of examples associated to the holomorphic-topological twist of $3d$ $\mathcal{N}=2$ theories was was discussed in \cite{Costello:2020ndc}, where it was argued in that the $3d$ bulk theory becomes topological when the boundary algebra contains a stress tensor. In this section, we prove a result along these lines for the more general $3d$ Poisson sigma model. 

Suppose that the Poisson vertex algebra $V$ contains an element $T \in V$ satisfying the Virasoro $\lambda$-bracket relation \be \{T \,_{\lambda} \, T\} = \frac{c}{12} \lambda^3 + 2 \lambda T + \del T\ee for some $c$. Let $T_{(n)} : V \rightarrow V$ be the operator defined by \be T_{(n)} v = \frac{\del^n}{\del \lambda^n} \{T \,_{\lambda} \, v \} |_{\lambda = 0}, \,\,\,\,\, v \in V,\,\,\,\, n  \in \{0,1,2, \dots,  \}.\ee We moreover suppose that $T_{(0)}$ acts on $V$ as the derivation $\del$, and $T_{(1)}$ acts semisimply on $V$ with (half)-integer eigenvalues. Recall that we encountered similar condition in Section \ref{sec:Global_defn} where we discuss $\mathrm{Der}(\mathbb{C}[z])$ action on $V$. Here the condition is stronger that we require the action to be inner. Our claim is that under these conditions, the holomorphic translations become a gauge symmetry, and the theory is therefore topological.

\begin{figure}[h!]
    \centering
    \begin{tikzcd}[row sep = 0.3em, column sep = 1.5em]
        \{\text{Poisson vertex algebra (PVA)}\} \arrow[r] & \{\text{Locally defined HT theory}\}\\
        \bigcup & \bigcup\\
        \{\text{PVA with a $\mathrm{Der}(\mathbb{C}[z])$ action} \} \arrow[r] & \{\text{Globally defined HT theory}\}\\
        \bigcup & \bigcup\\
        \{\text{PVA with an inner $\mathrm{Der}(\mathbb{C}[z])$ action} \} \arrow[r] & \{\text{Globally defined topological theory}\}\\
    \end{tikzcd}
    \caption{A hierarchy of Poisson vertex algebras and their corresponding Poisson sigma models.}
    \label{fig:PVA_HT_hierarchy}
\end{figure}
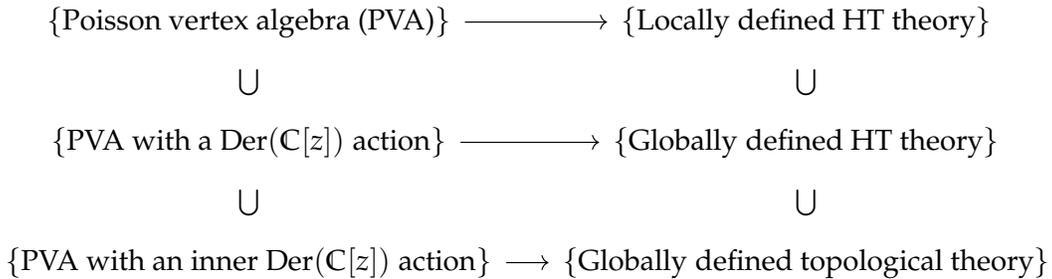

The proof is by considering the gauge transformation generated by $T$. By our assumption, $T$ acts on $\phi^i$ by
\begin{equation}
    \{T_{\lambda}\phi^i\}\big|_{\lambda = 0} = \partial\phi^i,
\end{equation}
We find that the transformation \be \delta_{T} = \{\int \text{d}z\, Q_{\mathrm{BV}}\mathbf{T},-\} \ee acts on the fields as \begin{align}
    &\delta_{T}\mathbf{\Phi}^i = -\del\mathbf{\Phi}^i\,,\\
    &\delta_{T}\boldsymbol{\eta}_i =  -\del\boldsymbol{\eta}_i\,.
\end{align}
As a consequence, the holomorphic translation $\partial$ is $Q_{\mathrm{BV}}$-exact and becomes a gauge symmetry, which implies that our theory is topological. \footnote{However, we do not know if the corresponding theory can be defined on any three-manifold or satisfies the complete axioms of a $3d$ TQFT. Nevertheless, the local operator algebra should be an $E_3$ algebra according to our argument.} 

 In the simplest example of the affine Kac-Moody algebra where \be \mathcal{V}(\mathfrak{g}) = \mathbb{C}[B_a, \del B_a, \dots],\ee the Virasoro element is given by the (classical) Sugawara construction: for $k \neq 0$ one has \be T = \frac{1}{2k} \kappa^{ab} B_a B_b,\ee which satisfies the Virasoro $\lambda$-bracket relations with $c = 0$. In this case, the holomorphic/topological theory can be written as Chern-Simons theory, which is manifestly topological. At critical level\footnote{As is well known and we will discuss later, the level is shifted by loop correction.}, which classically is $k=0$, the Poisson vertex algebra no longer contains the Virasoro element, and the $3d$ theory is not fully topological but only holomorphic-topological. 

In most other cases, the theory is only manifestly holomorphic-topological. The fact that the existence of a Virasoro element implies the fully topological properties becomes more intriguing. The passage from a holomorphic-topological theory to a fully topological theory typically requires extra supersymmetry and a further twist. However, our result predicts the existence of a large class of topological theories which do not come from the topological twist of any $3d$ supersymmetric theory.

\section{Deformation Quantization of Poisson Vertex Algebras} 
\label{sec:quant}
Having formulated our model and spelled out some examples, we now make some remarks on how the three-dimensional Poisson sigma model can potentially lead to a general strategy for the deformation quantization of Poisson vertex algebras to vertex algebras.

Given the holomorphic-topological Poisson sigma model associated to a Poisson vertex algebra, we formulate the model on a half-space $\mathbb{R}_+ \times \mathbb{C}$ with ``Neumann" boundary conditions at $t=0$. Supposing as before the superfields of the model are $\mathbf{\Phi}^i,\boldsymbol{\eta}_i$, the Neumann boundary conditions set $\boldsymbol{\eta}_i$ to vanish at the boundary, \be \boldsymbol{\eta}_i |_{t= 0 } = 0, \,\,\,\,i=1, \dots, n,\ee whereas $\mathbf{\Phi}^i$ are kept free. We are also left with a boundary BV differential $Q_{\text{BV}}^{\del}$ (where in the notation $\del$ is meant to stand for ``boundary" and not the holomorphic $z$-derivative)
\begin{equation}
    Q_{\text{BV}}^{\partial}\mathbf{\Phi}^i = (\text{d}_t + \ov{\partial})\mathbf{\Phi}^i\,.
\end{equation}

By passing to the cohomology, we see that the boundary algebra is generated by the fields $\phi^i$ and their $\partial_z$ derivatives. When the three-dimensional theory with Neumann boundary conditions can be successfully quantized, the algebra of local observables at the boundary, by virtue of being an algebra of observables supported along a holomorphic submanifold of a holomorphic-topological theory, is a holomorphic factorization algebra, namely a vertex algebra. The classical limit of this vertex algebra, as we will show below, is given by the original vertex Poisson algebra. Therefore, the full boundary vertex algebra, if it exists, is our proposed deformation quantization of the original Poisson vertex algebra. If the theory on $\mathbb{R}_+ \times \mathbb{C}$ has an obstruction to being quantized, we will say that the deformation quantization of the Poisson vertex algebra is obstructed. Perturbative quantization of more general holomorphic-topological theory is studied in \cite{Gwilliam:2021zkv,Wang:2024tjf}, and it is proved in \cite{Wang:2024tjf} that holomorphic-topological theory on $\mathbb{R}\times \mathbb{C}$ is unobstructed at odd-loop. However, we do not currently know if there are obstructions to quantizing the theory on $\mathbb{R}_+ \times \mathbb{C}$ at all loops.

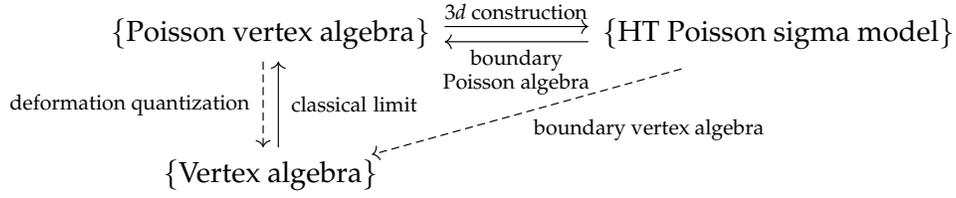
\begin{figure}[h!]
    \centering
    \begin{tikzcd}[row sep = 3em, column sep = 5em]
        \{\text{Poisson vertex algebra}\} \arrow[d,dashrightarrow,shift right,swap,"\text{deformation quantization}\;"] 
        \arrow[r,shift left, "3d \text{ construction}"] 
        & \{\text{HT Poisson sigma model}\} 
        \arrow[l,shift left,"\substack{\text{boundary}\\ \text{Poisson algebra}}"] 
        \arrow[dl,dashrightarrow,shift left,"\text{boundary vertex algebra}\;"]\\
        \{\text{Vertex algebra}\} 
        \arrow[u,swap,shift right,"\;\text{classical limit}"]& ~
    \end{tikzcd}
    \caption{Deformation quantization of PVA and quantization of the Poisson sigma model (with boundary).}
    \label{fig:PVA_HT_Chiral}
\end{figure}

We can summarize the construction in Figure \ref{fig:PVA_HT_Chiral}. 

\subsection{Boundary Poisson Vertex Algebras}
In this section, we analyze the boundary algebra to the first-order approximation. Though we do not know if the theory exists at the full quantum level, the tree-level Feynman diagrams are relatively simple and give rise to a well-defined "classical limit." To keep track of the order of deformation, it will be useful to rescale the interaction term in the action by a parameter$\hbar$:
\begin{equation}
    I = \sum_{n\geq 0}\frac{\hbar}{2} \boldsymbol{\eta}_i  \Pi_n^{ij}\partial^n \boldsymbol{\eta}_j \,.
\end{equation}

As we have mentioned, under the Neumann boundary condition, the boundary algebra is generated by the fields $\phi^i$ and their $\del_z$ derivatives. At order $\hbar^0$, we simply have the commutative algebra structure on $V = \IC[\partial^{n}\phi^{i},n\geq 0]$. Perturbative corrections to the OPE is computed via Feynman diagrams. The basic propagator, in the presence of boundary is given by \cite{Costello:2020ndc}
\begin{equation}
	\langle\mathbf{\Phi}(z,t)\boldsymbol{\eta}(z',t')\rangle = P_{\del}(z,t;z',t') = \frac{1}{2}(P(z,t;z',t') - P(z,t;z',-t'))\;,
\end{equation}
	where $P(z,t;z',t')$ is the bulk propagator
	\begin{equation}
		P(z,t;z',t') = \frac{1}{8\pi i}\frac{(\bar{z} - \bar{z}')d(t - t') - \frac{1}{2}(t - t')d(\bar{z} - \bar{z}')}{(|z - z'|^2 + (t - t')^2 )^{\frac{3}{2}}}\,.
	\end{equation}
The propagator itself is a one form, which means it pairs the 0-form component of $\mathbf{\Phi}$ with the 1-form component of $\boldsymbol{\eta}$. At order $\hbar$ there is only one Feynman diagram that contributes, given by the diagram appearing in the Figure below.  
\begin{figure}[h!]
	\begin{center}
		\begin{tikzpicture}[scale = 0.8]
		\draw[dashed](0,1) to (0,-1);
		\draw[dashed](0,1.8) to (0,2.2);
		\draw[dashed](0,-1.8) to (0,-2.2);
		\draw (0,1.4) circle [radius=0.4] node {$\phi^i$};
		\draw (0,-1.4) circle [radius=0.4] node {$\phi^j$};
		\draw (0.4,-1.4) arc (-90:-10:1.4);
		\draw (0.4,1.4) arc (90:10:1.4);
		\draw (1.8,0) circle [radius=0.2];
		\draw (2,0) to (3,0);
		\node [right,above] at (2,0) {$I$};
		\end{tikzpicture}
	\end{center}
    \caption{The tree level boundary Feynman diagram}
\end{figure}
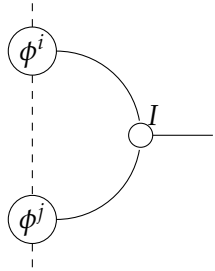

The above Feynman diagram gives us the OPE
\begin{equation}
    \phi^i(z)\phi^j(0) \sim \sum_{k\geq 0 }\int_{z' \in \IC,s\geq 0}(-1)^k\frac{\Pi^{ij}_k(\phi,\partial\phi,\dots)(s,z',\bar{z}')}{(|z - z'|^2 + s^2)^{\frac{3}{2}}}\partial_{z'}^k\frac{1}{(|z'|^2 + s^2)^{\frac{3}{2}}} \bar{z}s ds dz'd\bar{z}'\,.
\end{equation}
One can compute the above integral by Taylor expanding $\Pi^{ij}_k(s,z',\bar{z}')$ with respect to $s,\bar{z}',z'$. It is argued in \cite{Costello:2020ndc} that only $\Pi^{ij}_k(\phi,\partial\phi,\dots)(0)$ contribute to the singularity. Therefore, we only need to compute
\begin{equation}
    \mathcal{I}_k(z) = \int_{z' \in \IC,s\geq 0}\frac{1}{(|z - z'|^2 + s^2)^{\frac{3}{2}}}\partial_{z'}^k\frac{1}{(|z'|^2 + s^2)^{\frac{3}{2}}} \bar{z}s ds dz'd\bar{z}'\,.
\end{equation}
This integral is computed in \cite{Zeng:2023qqp}, giving us\footnote{We rescaled away an unimportant constant numerical factor here.}
\begin{equation}
    \mathcal{I}_k(z)  = \frac{(-1)^kk!}{z^{k+1}}\,.
\end{equation}
Altogether, we find the following OPE at order $\hbar$
\begin{equation}
    \phi^i(z)\phi^j(0) \sim \sum_{k\geq 0}\frac{k!}{z^{k+1}}\Pi^{ij}_k(\phi^i,\partial\phi^i,\dots)\,.
\end{equation}
The translation from an OPE into a \( \lambda \)-bracket is given by the Fourier transform \cite{Kac2015IntroductionTV}:
\begin{equation}
    a(z) \to \mathrm{Res}(e^{\lambda z}a(z))\,.
\end{equation}
We see that the pole \( \frac{k!}{z^{k+1}} \) contributes a \( \lambda^k \) to the \( \lambda \)-bracket. Therefore, the boundary algebra at first order gives us exactly the Poisson vertex algebra that we started with, as expected.

\subsection{Quantum Corrections}
\label{sec:q_correction}
\begin{figure}[h!]
	\begin{center}
		\begin{tikzpicture}[scale = 0.8]
		\draw[dashed](0,1.5) to (0,-1.5);
		\draw[dashed](0,2.3) to (0,2.7);
		\draw[dashed](0,-2.3) to (0,-2.7);
		\draw (0,1.9) circle [radius=0.4] node {$\phi^i$};
		\draw (0,-1.9) circle [radius=0.4] node {$\phi^j$};
		\draw (0.4,-1.9) arc (-90:-30:1.4);
		\draw (0.4,1.9) arc (90:30:1.4);
		\draw (1.65,1.1) circle [radius=0.1];
        \draw (1.65,-1.1) circle [radius=0.1];
        \draw (1.75,-1.1) arc (-70:70:1.15);
        \draw (1.55,1.1) arc (110:250:1.15);
		\end{tikzpicture}
	\end{center}
    \caption{The first loop digram}
    \label{loop}
\end{figure}
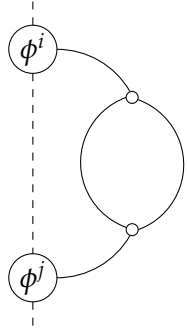

We now make some further remarks on loop corrections to the boundary vertex algebra. For example, the one loop effects typically involve Feynman diagrams such as the one occurring in Figure \ref{loop}.

While computing the corresponding Feynman integrals is expected to be challenging, one can obtain some non-trivial information by employing a familiar trick \cite{Gaiotto:2021tsq}. This involves considering the theory on $\mathbb{C} \times [0,L]$  and taking the length $L$ of the interval to vanish $L \to 0$. This results in a two-dimensional theory which is anomalous. Moreover, since the anomaly comes only from the boundaries and is symmetric, the anomaly of an individual boundary is half the anomaly of the resulting two-dimensional system. Since we can compute the anomaly of the two-dimensional system using standard conformal field theory techniques, this allows one to predict some of the quantum corrections to the classical Poisson vertex algebra without a direct Feynman diagram calculation.

We first illustrate this trick in the case of $BF$ theory. The theory that results upon slab reduction is the two-dimensional holomorphic field theory given by the gauged, adjoint valued $\beta\gamma$ system, \be S_{2d} = \int \beta_a \ov{\del} \gamma^a + f^a_{bc} A_{\ov{z}}^b \beta_a \gamma^c .\ee This is an anomalous system, and the resulting anomaly should be twice the value of the Feynman diagram responsible for the level shifting. This anomaly is known to be $-2 h^{\vee}$ and so this gives the level shifting effect \be k_{eff} = k-h^{\vee}\ee in the vertex algebra that quantizes the affine Poisson vertex algebra at level $k$.

We can employ a similar trick along with well-known results in conformal field theory to obtain the expected answers for some of our other examples. For the Virasoro theory, the theory we obtain by reduction on a slab has field content being a spin two-field $\beta$ and a spin $-1$ field $\gamma$. The general formula for the central charge of a spin $(s,1-s)$ is given by \cite{Polchinski:1998rq} \be c(s) = 3(2s-1)^2 -1\ee so we obtain $c = 26.$ Halving this we can obtain the quantum effective central charge \be c_{eff} = c+13.\ee Thus the deformation quantization of the classical Virasoro Poisson algebra with central charge $c$ is the quantum Virasoro vertex algebra with $c_{eff} = c + 13.$ For the superVirasoro algebra we obtain a similar result with 
\be c_{eff} = c + \frac{15}{2},\ee
since the theory obtained by slab reduction has $c= 15.$ These formulas are well known from bosonic and supersymmetric string theory. Finally the field content of the $W_3$-theory consists of a pair of $\beta\gamma$ systems, one with spin $(s,1-s)=(2,-1)$ and one with $(s',1-s')=(3,-2)$. This results in a system with central charge being \be c = (3(3)^2-1)+ (3(5)^2-1) = 100.\ee We expect the deformation quantization of the classical $W_3$ Poisson algebra to give the quantum $W_3$ vertex algebra with central charge 
\be c_{eff} = c+50.\ee

We leave it as an interesting challenge for the future to reproduce these formulas from Feynman diagrams of the holomorphic-topological theory. We expect that the methods developed in \cite{Budzik:2022mpd},\cite{Wang:2024tjf} can be applied to evaluate these numbers, and more generally, to produce the coefficients of the operator product expansion in the vertex algebra at the full quantum level.

\subsection{Koszul duality}
We also believe that the holomorphic-topological Poisson sigma model provides a natural framework to discuss Koszul duality for chiral algebras. This is a notion well established for associative algebras \cite{priddy1970koszul,beilinson1996koszul} and $E_n$ algebras \cite{ayala2014poincar}, but it is less studied for chiral algebras. For associative algebras, it is a functor that, roughly speaking, sends an algebra $A$ (over a field $k$) to $\mathrm{Ext}_{A}(k, k)$. This duality appears in the study of quantum field theories in various contexts, such as defects and boundaries \cite{Paquette:2021cij}. What is relevant for us is the expectation of finding Koszul duality for the boundary algebras associated with transverse boundary conditions \cite{Paquette:2021cij}.

So far, we have only considered the Neumann boundary condition, which should quantize the Poisson vertex algebra that we start with. We can also consider its transverse boundary condition, namely the Dirichlet boundary condition. This sets $\mathbf{\Phi}^i= 0$ at the boundary and leaves $\boldsymbol{\eta}_i$ free. As before, we can pass to the $(\text{d}_t + \bar{\del})$-cohomology and find that the boundary algebra is generated by the ghost fields $c_i(z)$ and their $\del_z$-derivatives. We conjecture that the two boundary algebras associated with the transverse boundary conditions (Dirichlet and Neumann) are Koszul dual to each other.

For an arbitrary $\lambda$-bracket, the boundary algebra of the Dirichlet boundary condition can be very complicated. We expect the notion of an ``$A_\infty$ chiral algebra" to arise in the most general scenario \cite{Costello:2020ndc}. Therefore, it will be instructive to analyze some simple examples where the vertex algebra structures for both boundary conditions are accessible.
\begin{itemize}
    \item  For a vanishing Poisson tensor, we get a free theory. Choosing Dirichlet boundary condition set $\mathbf{\Phi}^i = 0$. The boundary algebra is generated by the fields $\boldsymbol{\eta}_i$ with boundary BV differential
    \begin{equation}
        Q\boldsymbol{\eta}_i = (\text{d}_t + \bar{\del})\boldsymbol{\eta}_i\,.
    \end{equation}
 By passing to the cohomology, we see that the boundary algebra is generated by the ghost fields $c_i(z)$ and their derivatives. There is no boundary OPE so the boundary algebra is the graded commutative algebra $\mathbb{C}[c_i,\partial c_i,\dots]$, which is also a graded-commutative vertex algebra. This generalize the familiar example of the the Koszul duality between symmetric and exterior algebra.
 \item For a linear Poisson tensor defined by a Lie algebra $\mathfrak{g}$, we get the $3d$ holomorphic-topological BF theory. Under the Dirichlet boundary condition, and after passing to the $\text{d}_t + \bar{\del}$ cohomology, the boundary algebra is generated by the ghost fields $c_i(z),\partial c_i(z),\dots$. The boundary BV differential has the remaining action
 \begin{equation}
     Qc_i = \sum_{j,k} f_i^{jk}c_jc_k
 \end{equation}
 which can be identified with the Chevalley-Eilenberg differential of the Lie algebra $\mathfrak{g}[[z]]$. The interaction term does not introduce any OPE in the Dirichlet boundary condition. Therefore, we obtain a differential graded commutative algebra $(\mathbb{C}[c_i,\partial c_i,\dots],Q)$ as the boundary algebra. As we have discussed, the Neumann boundary condition gives rise to the affine Kac Moody VOA. This example generalizes the Koszul duality between the universal enveloping algebra and the Chevalley-Eilenberg algebra of a Lie algebra $\mathfrak{g}$.

 \item One can also consider the case of a quadratic Poisson tensor. In the case of the two-dimensional Poisson sigma model, a quadratic Poisson tensor is known to give rise to quadratic Koszul duality via deformation quantization \cite{shoikhet2010koszul}. We expect a similar story here. However, a general quadratic Poisson vertex algebra will introduce complicated loop corrections to the boundary algebra, which we cannot analyze yet. Nevertheless, we can construct a theory where the loop corrections vanish. The simplest such theory is given by a Poisson vertex algebra with four generators, where the only non-trivial $\lambda$-bracket is:
 \begin{equation}
     \{\phi^{1}~_\lambda \,\phi^2\} = \phi^3\phi^4\,.
 \end{equation}
 The corresponding Poisson sigma model has the standard kinetic term \eqref{eq:act_kin} and the following interaction term
 \begin{equation}
     \int\mathbf{\Phi}^1\mathbf{\Phi}^2\boldsymbol{\eta}_3\boldsymbol{\eta}_4\,.
 \end{equation}
In the Neumann boundary condition, we have the tree-level OPE 
\be 
\phi^1(z)\phi^2(0) \sim \frac{1}{z}\phi^3\phi^4(0), 
\ee without any loop corrections. In the Dirichlet boundary condition, there are also no loop corrections, and the OPE is given by \be c_3(z)c_4(0) \sim \frac{1}{z}c_1c_2(0) .\ee This reproduces an example found in \cite{Gui:2022pnx} for the quadratic Koszul duality of chiral algebras.
\end{itemize}

\section{Conclusion and outlook} \label{conc}

\paragraph{\textbf{Feynman diagrams and quantization}}
In this paper, we mainly considered the $3d$ Poisson sigma model at the classical level, only briefly studying its quantum properties. Understanding the full quantum theory is an important problem for the future. As we have explained, the problem of deformation quantization of PVAs relies on a consistent quantization of the $3d$ theory. However, at the moment, we do not know whether the $3d$ theory has higher loop anomalies. Understanding this requires a systematic analysis of the Feynman diagrams in the holomorphic topological theory. We believe that methods recently developed in \cite{Wang:2024tjf,Budzik:2022mpd} will be helpful in understanding these questions.

\paragraph{\textbf{VOA/Poisson sigma model correspondence}} The construction in this paper could be understood as a generalization of the famous Chern-Simons/WZW duality studied in \cite{Witten:1988hf}. As we have illustrated in our example, the Hilbert space of the theory admits a geometric construction from the moduli space of the equations of motion. This could have potential applications to the study of the conformal blocks of the boundary vertex algebra (if the quantization exists). We also briefly discussed Wilson lines in our examples. We believe incorporating them into the construction can reveal interesting connections with three-manifold and knot invariants.

\paragraph{\textbf{(Higher spin) gravity}} In our examples, we examined $3d$ models derived from the (super) Virasoro algebra and the $\mathcal{W}_3$ algebra, highlighting their connections to three-dimensional gravity as explored in \cite{Collier:2023fwi,Collier:2024mgv,Witten:1988hc}. A key direction for future research is to further investigate this connection, both at the classical and quantum levels. One advantage of our formulation is its natural extension to other $\mathcal{W}$ algebras. It would be interesting to study Poisson sigma models corresponding to other $\mathcal{W}_n$ algebras and $\mathcal{W}_\infty$ algebras, which may shed light on the study of three-dimensional higher spin gravity.

\paragraph{\textbf{Higher dimensional Poisson sigma models}}
Higher-dimensional holomorphic theories have gained much interest in recent studies \cite{Budzik:2023xbr,Gaiotto:2024gii}. The algebraic structure of local operators in such theories should be equipped with a higher-dimensional chiral algebra structure. The classical limit of the corresponding algebra should be described by some structure analogous to a Poisson vertex algebra. For example, a complex $n$-dimensional theory should correspond to a collection of $\lambda$-brackets involving $n$ $\lambda$-variables. Using a similar construction as in our paper, one can also construct higher-dimensional Poisson sigma models on $\mathbb{R} \times \mathbb{C}^n$. In addition, one can also try to construct fully holomorphic field theories starting from a (multivariable) $\lambda$-bracket. One such construction is given in Appendix \ref{fullyhol}.  It will be interesting to study these higher-dimensional models and their relationship with higher-dimensional chiral algebras.

\paragraph{\textbf{Relation to twisted holography}}
Twisted holography \cite{Costello:2018zrm} is a proposal to access SUSY-protected quantities on both sides of the well-known holographic dualities. A standard example is given in \cite{Costello:2018zrm}, as a twist of the famous $AdS_5/CFT_4$ duality \cite{Maldacena:1997re}. In this example, twisted holography is a duality between a large $N$ gauged $\beta\gamma$ VOA and the Kodaira-Spencer theory of gravity. In fact, it was shown that a proper dimensional reduction of the Kodaira-Spencer theory lands in our $3d$ Poisson sigma model \cite{Zeng:2023qqp}. Then, the twisted holography becomes an instance of the VOA/Poisson sigma model correspondence in our framework.

This construction becomes more useful if we consider the generalization of twisted holography studied in \cite{Gaiotto:2024dwr}. In that paper, the authors considered more general topological string theories, where a geometric description of the spacetime may not be available. In such cases, we no longer have a Kodaira-Spencer theory to describe the closed string sector. However, we should still be able to construct the dimensional reduction of the closed string field theory as one of our $3d$ Poisson sigma models.

\section*{Acknowledgments} We thank Kevin Costello, Davide Gaiotto, Nikita Sopenko and Mianghao Wang for useful discussions and comments. Some of this work was carried out while A.K. was at the Institute for Advanced Study. While there he was supported by the Bershadsky Fund and the Sivian Fund, along with the National Science Foundation under Grant No. PHY-2207584. A.K. and K.Z. are supported by the CMSA at Harvard University.

\appendix 

\section{Super Riemann Surfaces} \label{super}

In this appendix, we briefly review basic facts about super Riemann surfaces and their moduli. We refer to \cite{Witten:2012ga,Donagi:2013dua} for more detail.

A complex supermanifold is a locally ringed space $\mathfrak{X} = (X,\mathcal{O}_{\mathfrak{X}})$, such that $X$ is an ordinary complex manifold, $\mathcal{ O}_{\mathfrak{X}}$ is a sheaf of commutative complex super algebras and it is locally isomorphic to $\mathcal{O}_X\otimes \wedge^{\bullet}E$ for some vector space $E$. We denote $\mathfrak{X}_{\text{red}} = (X,\mathcal{O}_X)$ the ordinary complex manifold underling $\mathfrak{X}$. Given a pair $(X,\mathcal{E})$ with $X$ a complex manifold and $\mathcal{E}$ a vector bundle on it, one can construct supermanifold $\mathfrak{X}$ by letting $\mathcal{O}_{\mathfrak{X}} = \wedge^{\bullet}\mathcal{E}$. A supermanifold is said to split if it is isomorphic to one of the form $(X,\wedge^{\bullet}\mathcal{E})$. A supermanifold of dimension $n|1$ always split.

A super Riemann surface is a a complex supermanifold $\mathfrak{X}$ of dimension $1|1$, together with a subbundle $\mathcal{D}\subset T_{\mathfrak{X}}$ of rank $0|1$ which is required to be “completely nonintegrable”. This means that for any odd vector field $v \in \mathcal{D}$, its square $v^2 = \frac{1}{2}\{v,v\}$ is independent of $v$. This property can also be represented as the short exact sequence
\begin{equation}\label{eqn:SConf_exact}
    0 \to \mathcal{D} \to T_{\mathfrak{X}} \to \mathcal{D}^2\to 0\,.
\end{equation}
Such a distribution $\mathcal{D}$ is also called a superconformal structure $\mathfrak{X}$. In a local coordinate $(z,\theta)$, we have a superconformal vector $D_{\theta} \in \mathcal{D}$ 
\begin{equation}
    D_{\theta} = \frac{\partial}{\partial \theta} + \theta\frac{\partial}{\partial z}\,.
\end{equation}
We see that $D_{\theta}^2 = \frac{\partial}{\partial z}$, which indeed is independent of $D_{\theta}$. In fact, locally on a super Riemann surface we can always choose a coordinate such that $\mathcal{D}$ is generated by the above vector $D_{\theta}$ \cite{Donagi:2013dua}.

As a $1|1$ dimensional supermanifold, a super Riemann surface always split. Therefore it is specified by a ordinary Riemann surface $\Sigma$ and a line bundle $\mathcal{E}$. In fact, the superconformal structure fix the line bundle to satisfy $\mathcal{E}^{\otimes 2} = K_{\Sigma}$, so $\mathcal{E}$ is a spin structure on $\Sigma$. Therefore, a super Riemann surface has the same data as a spin curve.

\subsection{Super Moduli Space}
\label{app:Super_RS}
An important structure of a super Riemann surface is the subbundle $\mathcal{S} \subset T_{\mathfrak{X}}$ that preserve the superconformal structure $\mathcal{D}$. A vector field $v \in \mathcal{S}$ is called a superconformal vector field. In a local coordinate, they are the vector fields whose commutator with $D_{\theta}$ is still propotional to $D_{\theta}$. We can check that $\mathcal{S}$ has a basis of even vector fields
\begin{equation}
    a(x)\frac{\partial}{\partial z} + \frac{\partial_za(z)}{2}\theta\frac{\partial}{\partial \theta}
\end{equation}
and odd vector fields
\begin{equation}
    b(x)(\frac{\partial}{\partial \theta} - \theta\frac{\partial}{\partial z})\,.
\end{equation}
In fact, the even and odd part can be combined to give
\begin{equation}
    f(z,\theta)\frac{\partial}{\partial z} + \frac{D_{\theta}f(z,\theta)}{2}D_{\theta}\,,
\end{equation}
where $f(z,\theta) = a(z) + 2\theta b(z) $. This also implies that $\mathcal{S} = T_{\mathfrak{X}}/\mathcal{D} \cong \mathcal{D}^{2}$.

A superconformal vector field is a infinitesimal version of a superconformal transformation, which is a coordinate transformation $(z,\theta) \to (\tilde{z},\tilde{\theta})$ that preserve the bundle $\mathcal{D}$. We can check that a general coordinate transformation acts on the vector $D_{\theta}$ by
\begin{equation}
    D_{\theta}\tilde{\theta}D_{\tilde{\theta}} + (D_{\theta}\tilde{z} - \tilde{\theta}D_{\theta}\tilde{\theta})D_{\tilde{\theta}}^2\,.
\end{equation}
Therefore, a transformation $(z,\theta) \to (\tilde{z},\tilde{\theta})$ is superconformal if and only if it satisfy
\begin{equation}\label{eqn:sconf_trans}
    (D_{\theta}\tilde{z} - \tilde{\theta}D_{\theta}\tilde{\theta})\,.
\end{equation}

Let us denote $\mathfrak{M}_g$ the moduli space of super Riemann surface of genus $g$. Analogous to the non-super case, the tangent space of $\mathfrak{M}_g$ is given by $T\mathfrak{M}_g|_{\mathfrak{X}} = H^{1}(\mathfrak{X},\mathcal{S})$. It consist of an even and odd subspaces denoted $T_{\pm}\mathfrak{M}_g|_{\mathfrak{X}}$. According to ou previous discussion, we have
\begin{align}
    T_{+}\mathfrak{M}_g|_{\mathfrak{X}} &= H^{1}(\Sigma,T_{\Sigma})\,,\\
    T_{-}\mathfrak{M}_g|_{\mathfrak{X}} &= \Pi H^{1}(\Sigma,T_{\Sigma}^{\frac{1}{2}})\,.
\end{align}

Another important structure of a super Riemann surface is the canonical bundle, which is the bundle of holomorphic volume form. Recall that the canonical bundle for an ordinary complex manifold is the determinant of its cotangent bundle. For a complex supermanifold, its canonical bundle is defined as the Berezinian of its cotangent bundle $\mathrm{Ber}_{\mathfrak{X}} = \mathrm{Ber}(T^*_{\mathfrak{X}})$. To compute it we take the dual of \eqref{eqn:SConf_exact}
\begin{equation}
    0 \to \mathcal{D}^{-1} \to T^*_{\mathfrak{X}} \to \mathcal{D}^{-2}\to 0\,.
\end{equation}
This implies that \be \mathrm{Ber}_{\mathfrak{X}} \cong \mathrm{Ber}(\mathcal{D}^{-1})\otimes \mathrm{Ber}(\mathcal{D}^{-2}) .\ee Berezinian has the property that it send a even number $a$ to $\mathrm{Ber}(a) = a$ and an odd number $b$ to $\mathrm{Ber}(b) = b^{-1}$. From our previous computation we see that $\mathcal{D}^{2}$ is a rank $1|0$ even bundle and $\mathcal{D}$ is a rank $0|1$ odd bundle. Therefore, we find that
\begin{equation}
    \mathrm{Ber}_{\mathfrak{X}} \cong \mathcal{D} \otimes \mathcal{D}^{-2} = \mathcal{D}^{-1}\,.
\end{equation}
We also have a super version of Serre duality, as a non-degenerate pairing
\begin{equation}
    H^{k}(\mathfrak{X},\mathcal{E})\otimes \Pi H^{1-k}(\mathfrak{X},\mathcal{E}^{*}\otimes \mathrm{Ber}_{\mathfrak{X}}) \to \mathbb{C}\,.
\end{equation}
This allows us to identify the cotangent space $T^*\mathfrak{M}_g|_{\mathfrak{X}}$ of the moduli space as
\begin{equation}
    T^*\mathfrak{M}_g|_{\mathfrak{X}} = H^{1}(\mathfrak{X},\mathcal{S})^*  = \Pi H^{0}(\mathfrak{X},\mathcal{D}^{-3})\,.
\end{equation}
We can identify its even and odd part to be
\begin{align}
        T_{+}^*\mathfrak{M}_g|_{\mathfrak{X}} &=H^{0}(\Sigma,K^{2})\,,\\
    T_{-}^*\mathfrak{M}_g|_{\mathfrak{X}} &=\Pi H^{0}(\Sigma,K^{\frac{3}{2}})\,.
\end{align}

\subsection{Super Projective Structure}
\label{app:Super_Proj}
In this section, we introduce the definition of super projective connections and their relations to super projective structures and opers. We refer to \cite{Zeitlin:2013iya} for more detail.

A super projective connection is a differential operator of $\mathcal{D}^{-1} \to \mathcal{D}^2$, such that in a local coordinate, it take the following form
\begin{equation}
    D_{\theta}^3 + \omega(z,\theta)\,.
\end{equation}
Under a superconformal coordinate transformation $(z,\theta) \to (\tilde{z},\tilde{\theta})$, $\omega(z,\theta)$ transform as
\begin{equation}
    \tilde{\omega}(\tilde{z},\tilde{\theta}) = (D_{\theta}\tilde{\theta})^{3}\omega(\tilde{z},\tilde{\theta}) + \{\tilde{\theta};z,\theta\},
\end{equation}
where $\{\tilde{\theta};z,\theta\}$ is the super Schwarzian derivative \cite{Friedan:1986rx}:
\begin{equation}
    \{\tilde{\theta};z,\theta\} = \frac{\partial_z^2\tilde{\theta}}{D_{\theta}\tilde{\theta}} - 2\frac{\partial_z\tilde{\theta}D_{\theta}^3\tilde{\theta}}{(D_{\theta}\tilde{\theta})^2}\,.
\end{equation}
We consider a infinitesimal coordinate transformation $\tilde{z} = z + v,\tilde{\theta} = \theta + \gamma$. Then \eqref{eqn:sconf_trans} implies that $\gamma = \frac{1}{2}D_{\theta}v$. We find that $D_{\theta}\tilde{\theta} \approx \frac{1}{2}\partial_z v $ and
\begin{equation}
    \{\tilde{\theta};z,\theta\} \approx \frac{1}{2}D_{\theta}\partial_z^2v\,.
\end{equation}
Therefore, we see that the gauge transformation \ref{eqn:gauge_super} for $\frac{6}{c}\mathfrak{T} = \frac{6}{c}(G + \theta T)$ is indeed the infinitesimal transformation of super projective connection. 

Super projective connection is closely related to super projective structure. By definition, a super projective structure on a super Riemann surface is an equivalence class of coverings, where the transition functions between overlaps are given by super M\"{o}bius transformations.
\begin{align}
    \begin{split}
        z &\to \frac{az+b}{cz+d} + \theta\frac{ez + f}{(cz+d)^2}\,,\\
        \theta &\to \frac{ez + f}{cz+d} + \theta\frac{1 + \frac{1}{2}ef}{cz+d}\,,
    \end{split}
\end{align}
where $a,b,c,d$ are even variables with $ac-bd = 1$ and $e,f$ are odd variables.
It is shown in \cite{Zeitlin:2013iya} that there is a bijection between the set of super projective structures and the set of super projective connections on a super Riemann surface. The basic idea is that a super projective coordinate chart can be constructed from the solutions of $(D_{\theta}^3 + \omega(z, \theta))\psi = 0$ out of a super projective structure.

\cite{Zeitlin:2013iya} further proves that there is a bijection between super projective connections and $\mathfrak{osp}(2|1)$ opers. This bijection corresponds to the following $\mathfrak{osp}(2|1)$ super connection:\begin{equation}
    D_{\theta} + \begin{pmatrix}
        0&0&\omega(z,\theta)\\-1&0&0\\0&1&0
    \end{pmatrix}\,.
\end{equation}

\section{Fully Holomorphic Poisson Sigma Model} \label{fullyhol}

In this appendix we briefly mention a fully holomorphic four-dimensional variant of the three-dimensional Poisson sigma model. The fully holomorphic Poisson sigma model is a four-dimensional field theory formulated on $\mathbb{C}^2$ with standard holomorphic coordinates $(z_1, z_2)$. The fundamental fields of the model consist of $N$ scalar fields \be \phi^i(z_1,\bar{z}_1, z_2, \bar{z}_2), \,\,\,\,\, i = 1, \dots, N, \ee along with $N$ anti-holomorphic one-forms \be \eta_i = \eta_{i\ov{z_1}} \text{d} \ov{z_1} + \eta_{i\ov{z_2}}\text{d}\ov{z_2}, \,\,\,\,\, i = 1, \dots, N.\ee We can now consider a differential operator of the form \be \Pi^{ij}(\del_{1}, \del_{2}) = \sum_{n,m \geq 0} \Pi^{ij}_{n,m}(\phi, \del_a \phi, \del_a \del_b \phi, \dots) (\del_1)^n (\del_2)^m \ee which is assumed to be skew-hermitian as before: \be -\Pi^{ij}(\del_1, \del_2) = \sum_{n,m} (-1)^{n+m}\del_1^n \del_2^m  \circ \Pi^{ji}_{n,m}(\phi,\del_a \phi, \del_a\del_b \phi, \dots) =(\Pi(\del_1, \del_2)^{\dagger})^{ij}.\ee Letting $\ov{\del}$ be the Dolbeault operator on $\mathbb{C}^2$, which for instance acts on scalars $\phi$ via
\be
\ov{\del} \phi = \frac{\del \phi}{\del \ov{z}_1} \text{d}\ov{z}_1 + \frac{\del \phi} {\del \ov{z}_2} \text{d}\ov{z}_2 
\ee the action of the four-dimensional Poisson sigma model reads \be S = \int_{\mathbb{C}^2} \text{d}z \text{d}w \Big( \eta_i \ov{\del} \phi^i + \frac{1}{2} \eta_i \,\Pi^{ij}(\del_1, \del_2) \eta_j \Big). \ee The fields are subject to the gauge transformation rules written in BRST form 
\begin{align}
Q \phi^i &= \Pi^{ij}(\del_1, \del_2) c_j, \\ 
Q  \eta_i &= -\ov{\del} c_i -\frac{1}{2} \sum_{n,m} (-1)^{n+m} \del_1^n \del_2^m \Big( \eta_j \frac{\del \Pi^{jk}(\del_1, \del_2)}{ \del (\del_1^n \del_2^m \phi^i)} c_i - c_j \frac{\del \Pi^{jk}(\del_1, \del_2)}{ \del (\del_1^n \del_2^m \phi^i)} \eta_i \Big), \\ Q c_i &= -\frac{1}{2} \sum_{n,m \geq 0 } (-1)^{n+m}\del_1^n \del_2^m \Big( c_j\frac{ \del \Pi^{jk}}{\del(\del_1^n \del_2^m \phi^i)} c_k \Big).
\end{align} The model written as above will possess the above gauge invariance and \be Q^2 =0  \ee  will hold on-shell if and only if the identity
\begin{align} 
\begin{split} 
\sum_{n,m \geq 0} \Big(\frac{\del \Pi^{kj}(\del_1, \del_2)}{ \del (\del_1^n \del_2^m \phi^l)} G_j\Big)  \big( \del_1^n \del_2^m \big (\Pi^{li}(\del) F_i \big) \big) - \Big( \frac{\del \Pi^{ki}(\del)}{\del(\del_1^n \del_2^m \phi^l)} F_i \Big) (\del_1^n \del_2^m (\Pi^{lj}(\del_1, \del_2) G_j) \\ = \sum_{n \geq 0 } \Pi^{kl}(\del_1, \del_2) (-1)^{n+m}\del_1^n \del_2^m \Big(G_j \frac{\del \Pi^{ji}(\del)}{ \del (\del_1^n \del_2^m \phi^l)} F_i \Big),
\end{split} 
\end{align} holds. This is equivalent to saying that  \be \Pi^{ij}(\lambda_1, \lambda_2) = \sum_{n,m \geq 0} \Pi^{ij}_{n,m} (\phi, \del_a \phi, \dots) \lambda_1^n \lambda_2^m\ee defines a two-variable $\lambda$-bracket \be \{\phi^i \,_{(\lambda_1, \lambda_2)} \, \phi^j\} = \Pi^{ji}(\lambda_1, \lambda_2)\ee satisfying the two-variable Jacobi identity and thus equips the space \be V = \mathbb{C}[\phi, \del_a \phi, \del_a \del_b \phi, \dots] , \,\,\,\, a,b \in \{1,2\},\ee with the structure of a two-variable Poisson vertex algebra. 
\begin{align}
\sum_{n \geq 0} \frac{\partial \Pi^{kj}(\mu)}{\partial (\partial_1^n \partial_2^m \phi^l)} (\partial_1 + \lambda_1)^n (\partial_2 + \lambda_2)^m \Pi^{li}(\lambda)
- \sum_{n \geq 0} \frac{\partial \Pi^{ki}(\lambda)}{\partial (\partial_1^n \partial_2^m \phi^l)} (\partial_1 + \mu_1)^n (\partial_2 + \mu_2)^m \Pi^{lj}(\mu) \nonumber \\
= \sum_{n \geq 0} \Pi^{kl}(\lambda + \mu + \partial) \big(-(\lambda + \mu + \partial)\big)^n \frac{\partial \Pi^{ji}(\lambda)}{\partial (\partial^n \phi^l)}.
\end{align}

One of the simplest example of the four-dimensional Poisson sigma model is given by letting $V$ be \be V = \text{Sym}(\mathbb{C}[\del_1, \del_2]\otimes \mathfrak{g}) \ee for a Lie algebra $\mathfrak{g}$ equipped with a Killing form $\kappa_{ab}$.  The $(\lambda_1, \lambda_2)$-bracket on generators $\{B_a \}$ is given by \be \{B_a \,_{(\lambda_1, \lambda_2)} B_b \} = f_{ab}^c B_c.\ee The corresponding theory is the purely holomorphic BF theory in four-dimensions, which arises as the holomorphic twist of pure $\mathcal{N}=1$  supersymmetric Yang-Mills theory. This was studied recently in \cite{Budzik:2023xbr}. We can also deform the $(\lambda_1, \lambda_2)$-bracket to be \be \{B_a \,_{(\lambda_1, \lambda_2)} B_b \} = f_{ab}^c B_c + \lambda_1 k \kappa_{ab},\ee and the corresponding theory becomes, after a field redefinition, the holomorphic-topological four-dimensional Chern-Simons theory \cite{Costello:2013zra}. Choosing a direction in the $(\lambda_1, \lambda_2)$-space, say $\lambda_1$ as above, corresponds to a choice of a topological plane $\mathbb{R}^2$ in the four-dimensional spacetime $\mathbb{C}^2$.

At this point it is natural to wonder if there are other dimensions in which analogous mixed holomorphic-topological Poisson sigma models exist. There is a simple counting argument which excludes this once we assume a certain form of the fields. Suppose the space-time takes the form $X_{(d,k)} := \mathbb{C}^d \times \mathbb{R}^k$ and we let $\eta$ be a one-form valued in an $N$-dimensional vector space $L$ with anti-holomorphic directions only along $\mathbb{C}^d$ and $\phi: X_{(d,k)} \to L$. The natural Lagrangian of the purported model is \be  L_{(d,k)} = \text{d}z^1\dots \text{d}z^d \Big(\eta_i(\text{d}_{\mathbb{R}^k} + \ov{\del}_{\mathbb{C}^d}) \phi^i + \frac{1}{2}\eta_i\Pi^{ij}(\del_1, \dots, \del_d) \eta_j \Big),\ee which is an $(n+2)$-form on $X_{(d,k)}$. In order to integrate it over $X_{(d,k)}$ to obtain a well-defined action we must have \be d+2 = 2d+k \implies d= 2-k \ee so that the only solutions are \be (d,k)= (0,2), \,\, (1,1),  \,\, (2,0) .\ee These correspond to the two, three and four-dimensional Poisson sigma models respectively. It is worth nothing that a similar argument tells us that the only mixed holomorphic-topological Chern-Simons theories (where the field is a connection one-form $A$) are in three, four, five and six dimensions.

We will not pursue developing the fully holomorphic variant of Poisson sigma model any further in this note. It remains a fascinating theory that warrants further study.

\end{document}